 \documentclass[final,5p,times,twocolumn]{elsarticle}

\usepackage{amssymb}
\usepackage{url}
\usepackage{soul}
\usepackage{color}
\usepackage{graphics}
\usepackage{graphicx}
\usepackage{multirow}
\graphicspath{{}}
\journal{Journal of Web Semantics}

\begin{document}

\begin{frontmatter}

\title{A Hybrid Approach to Finding Relevant Social Media Content for Complex Domain Specific Information Needs}
\address[label1]{Ohio Center of Excellence in Knowledge-enabled Computing (Kno.e.sis)\\Wright State University, Dayton OH 45435, USA}

\author{Delroy Cameron}
\ead{delroy@knoesis.org}
\cortext[cor1]{Corresponding Author. Tel.: +1 937 775 5213; fax: +1 937 775 5133}	

\author{Amit P. Sheth}	

\author{Nishita Jaykumar}

\author{Krishnaprasad Thirunarayan}

\author{Gaurish Anand}

\author{Gary A. Smith}

\begin{abstract}
While contemporary semantic search systems offer to improve classical keyword-based search, they are not always adequate for complex domain specific information needs. The domain of prescription drug abuse, for example, requires knowledge of both ontological concepts and ``intelligible constructs'' not typically modeled in ontologies. These intelligible constructs convey essential information that include notions of intensity, frequency, interval, dosage and sentiments,  which could be important to the holistic needs of the information seeker. In this paper, we present a hybrid approach to domain specific information retrieval (or knowledge-aware search) that integrates ontology-driven query interpretation with synonym-based query expansion and domain specific rules, to facilitate search. Our framework is based on a context-free grammar (CFG) that defines the query language of constructs interpretable by the search system. The grammar provides two levels of semantic interpretation: 1) a top-level CFG that facilitates retrieval of diverse textual patterns, which belong to broad templates and 2) a low-level CFG  that enables interpretation of certain specific expressions that belong to such patterns. These low-level expressions occur as concepts from four different categories of data: 1) ontological concepts, 2) concepts in lexicons (such as emotions and sentiments), 3) concepts in lexicons with only partial ontology representation, called \textit{lexico-ontology} concepts (such as side effects and routes of administration (ROA)), and 4) domain specific expressions  (such as date, time, interval, frequency and dosage) derived solely through rules. Our approach is embodied in a novel Semantic Web platform called PREDOSE, which provides search support for complex domain specific information needs in prescription drug abuse epidemiology. When applied to a corpus of over 1 million drug abuse-related web forum posts, our search framework proved effective in retrieving relevant documents when compared with three existing search systems.
\end{abstract}

\begin{keyword}
Semantic Search \sep Information Retrieval \sep Complex Information Needs \sep Ontology \sep Background Knowledge \sep Context-Free Grammar, Knowledge-Aware Search



\end{keyword}

\end{frontmatter}


\section{Introduction}\label{intro}
The use of structured background knowledge (ontologies) to enhance search  has gained considerable traction among contemporary information retrieval systems. Ontologies offer to  improve search by capturing the meaning of real-world concepts and their associations.  The formal representations modeled in ontologies have been used to positively impact many complex tasks, including interoperability, personalization and knowledge discovery. 

While semantic search has gained credibility, compared to classical keyword-based and hyperlinked-based search, there is often a misalignment between the information needs of users and the knowledge model developed to meet such needs. Ontologies provide a means for interpreting some elements of complex information needs, but not all aspects of such needs \cite{FernandezCLVCM11}. The main issue is that ontologies often have limited scope, while users are unrestricted in the range of information they can seek on a given topic. A user information need can transcend data types and sources, exceeding what is formally modeled.

In spite of this, many semantic search applications \cite{TranCRS07, LeiUM06, GuhaMM03, Rocha2004}, semantic search engines (Hakia, Bing) and hybrid information retrieval approaches \cite{FernandezCLVCM11, ValletFC05, CastellsFV07, Ruotsalo12} rely heavily on ontologies for query interpretation. While these approaches serve their intended purpose, they are generally unsuitable for  domain specific applications, such as prescription drug abuse. General-purpose search engines such as Google and Yahoo that rely on keyword-based and hyperlinked-based models, may not perform well on domain specific data. This is because minimal (and often inadequate) support is provided for interpreting the additional elements that could be important to an information need, but not formally captured by the knowledge model. 

We address this problem by developing and evaluating a hybrid approach to search that allows query specification and interpretation of diverse expressions, which are involved in various aspects of complex information needs. To illustrate our approach, consider a scenario in which an epidemiologist in the domain of prescription drug abuse is seeking insights into emerging patterns and trends in drug abuse using social media. For brevity, we  present only one of many information needs explored in PREDOSE  (\url{http://wiki.knoesis.org/index.php/PREDOSE}). \\

\noindent \textit{\textbf{Information Need:} ÔHow are drug users engaging in the use of the semi-synthetic opioid Buprenorphine, through excessive daily dosage?Õ} \\

\noindent Inherent in this information need is the following relevant background knowledge. Buprenorphine is an opioid antagonist used in the treatment of opioid addiction, including addiction to Heroin, OxyContin and Vicodin. Prescribed daily dosage varies by individual ranging from 4--32mg\footnote{Note that the actual amounts used in examples throughout this manuscript are anecdotal only}. Buprenorphine is known to stabilize drug users from withdrawal symptoms, but can also induce an opiated effect. This treatment drug is therefore at risk for abuse. Epidemiologists are interested in understanding the dosage practices of Buprenorphine users, including amounts taken, frequency of use and side effect experienced, to better understand emerging patterns and  trends of abuse. 

A suitable user query provided by a domain expert may involve the following keywords: ``buprenorphine dosage exceed 4mg daily.'' A robust search system may correctly interpret the keyword `buprenorphine' as the standard DBpedia resource: http://dbpedia.org/resource/Buprenorphine. Then through non-trivial query expansion, the system may also associate the keywords `bupe,' `bupey,' `suboxone,' `subbies' and `suboxone film,' with `Buprenorphine,' as synonyms. Likewise, the search system may expand the keyword `daily' with the synonyms `day,' `night,' `morning' and `afternoon,' using available (or manually created) lexicons that contain such mappings. However, the intricate challenge is interpreting the notion of Ôexcessive dosage,Õ expressed as the phrase ``dosage exceed 4mg.'' 

In the development of Active Semantic Electronic Medical Records (ASEMR), Sheth et. al. \cite{ShethALOWYG07} created rules expressed in RDQL \cite{Seaborne04} (precursor to SPARQL) to enable specification of additional constructs (including dosage) that compensate for deficiencies in the knowledge model. Similarly, in the Semantic Content Organization and Retrieval Engine (SCORE) \cite{Hammond02, sheth2001system}, Hammond et. al. implemented various rules derived using regular expressions to specify quantity-conveying metadata (such as `currency,' `percentage,' `amount,' `time' and `dates') which were not present in the ontology. In  the Knowledge and Information Management platform (KIM) \cite{PopovKOMK04}, Popov et. al. modeled various lexical resources in the ontology such as currency, dates and abbreviations, which were subsequently used for document annotation. However, the information need presented here requires a more in-depth interpretation. 

To appropriately interpret excessive dosage, the notion of dosage itself must first be specified using its constituent members: {\footnotesize DOSAGE-OPERATOR} (e.g., `$>$,' `$<$'), {\footnotesize DOSAGE-AMOUNT} (e.g., `4,' `10') and {\footnotesize DOSAGE-UNIT} (e.g., `mg,' `tablet'). In this way, the search term `$>$4mg'  could be an abstraction of the search phrase ``dosage exceed 4mg.'' Rules must then be used to interpret each constituent according to what is possible in the corpus. This is important because a {\footnotesize DOSAGE-UNIT}  may have various lexical representations in text (e.g., Ômg,Õ Ômilligram,Õ Ômilli-gramÕ). Likewise, the {\footnotesize DOSAGE-OPERATOR} can have multiple equivalent manifestations (such as `$>$', `greater than,' `more than' and `above'). Similarly, {\footnotesize DOSAGE-AMOUNT}  can be numeric or textual. According to these possible interpretations, `6mg,' `ten milligrams,' `about 8mgs,' `a bit more than 30 milli-grams' etc, are all valid expressions for the query `dosage exceed 4mg.' The matching documents for the entire query (``buprenorphine dosage exceed 4mg daily'') are obtained after filtering heuristics are applied to retrieve text fragments from the corpus that match the interpretation of each query component. In this way, a hybrid approach to information retrieval would have been utilized, which leverages ontologies, lexicons and rules for query interpretation of domain specific data.

Concretely, our approach is based on a context-free grammar (CFG) that defines the query language of constructs interpretable by the search system. The grammar provides two levels of semantic interpretation: 1) a top-level CFG defines broad patterns  that can be interpreted by the system, and 2) a low-level CFG     defines the interpretation of certain specific elements that belong to such patterns. The query language of the grammar is specified in an IBM declarative information extraction specification called SystemT \cite{Chiticariu2010, KrishnamurthyLRRVZ08}, which is designed for information extraction from heterogeneous texts. SystemT is advantageous because it enables porting of rules to texts in other domains. Some of these domains specifically: 1) biomaterials and materials science, 2) cannabis and synthetic cannabinoid research and 3) clinical texts on cardiology reports. 

In an evaluation using a corpus of over 1 million web forum posts related to drug abuse, our hybrid search system retrieved a larger number of relevant documents when compared with three existing search systems. These systems are the: 1)  semantic search engine Hakia, 2)  crowd sourcing-based search engine DuckDuckGo  and 3)  popular search engine Google. Note that since these search engines are not specifically engineered to handle domain specific data,  our results are not surprising. However, our experiments highlight the need for more effective approaches to domain specific search as noted in \cite{Broder2002}. The specific contributions of this research are as follows:
\begin{itemize}
	\item We develop a hybrid approach to domain specific information retrieval that interprets four categories of data. These are: 1) structured background knowledge in ontologies, 2) concepts in lexicons, 3) concepts in lexicons with partial ontology representation called \textit{lexico-ontology} concepts (see Section \ref{knomo}), and 4) concepts defined using rules.
	\item We utilize a CFG to formally define the query language of strings interpretable by the system. The CFG provides two levels of semantic interpretation: 1) a top-level CFG for interpreting general textual patterns, and 2) a low-level CFG for interpreting specific expressions.
	\item We show that our approach is effective through an evaluation against three popular search systems.
\end{itemize}

The rest of the paper is organized as follows. Section \ref{framework} describes the overall hybrid information retrieval framework, which includes modules for query interpretation in Section \ref{qinterpret}, semantic metadata extraction/document annotation in Section \ref{docannot} and query matching in Section \ref{qmatch}. Section \ref{eval} describes the evaluation  and Section \ref{related} covers related work. 

\section{Approach}\label{framework}
\begin{figure}[htp]
	\centering
		\caption{System Architecture}\label{fig:architecture}
		\includegraphics[scale=0.51]{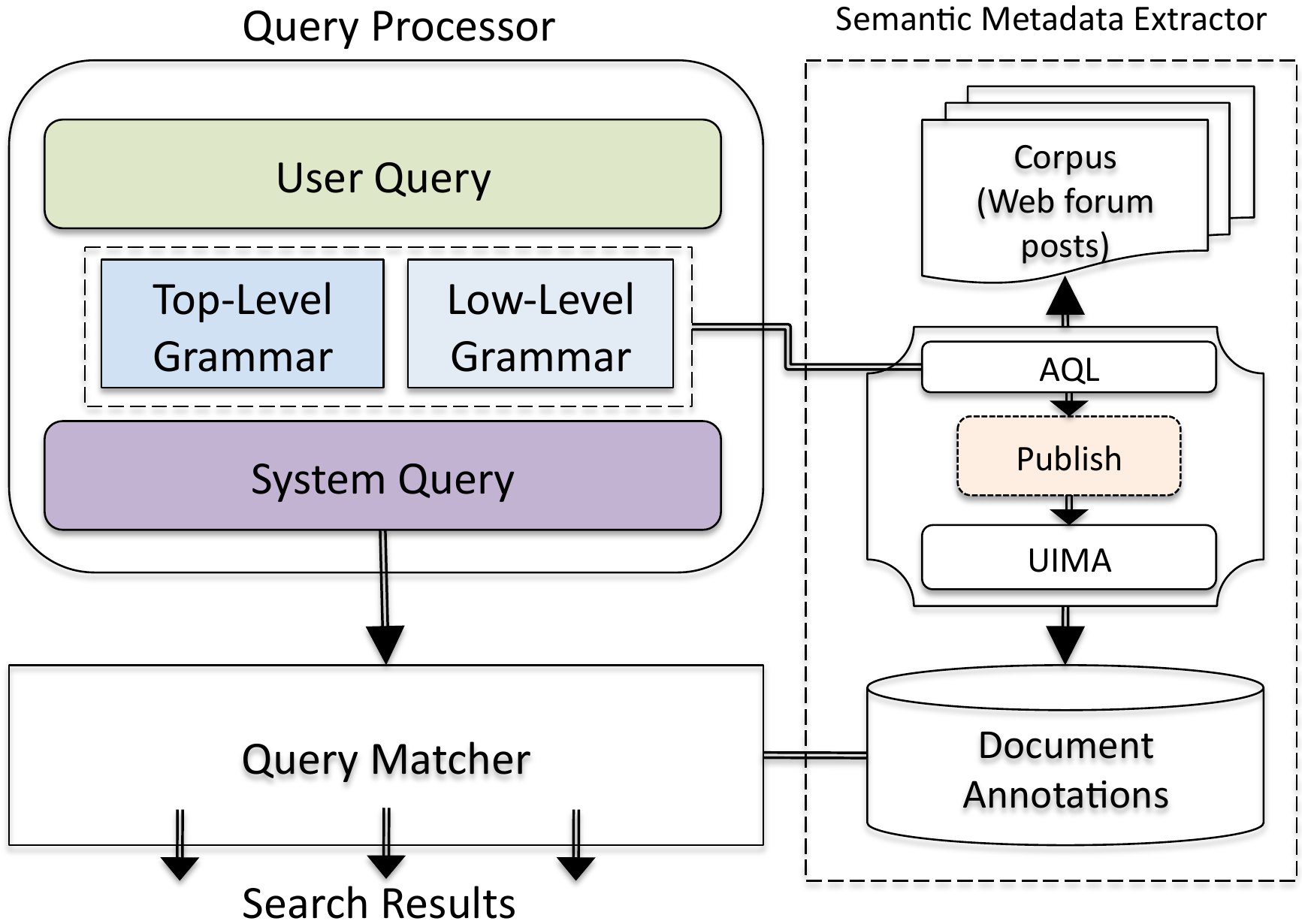}
\end{figure}

Our hybrid information retrieval system (shown in Figure \ref{fig:architecture}) consists of three components: 1) Query Processor, 2) Semantic Metadata Extractor and 3) Query Matcher. The \textit{query processor} provides functionality for template-based query specification and domain specific query interpretation. The \textit{semantic metadata extractor} identifies the offsets of text snippets that match the query interpretation in the corpus. The \textit{query matcher} retrieves and filters the relevant documents for a given user query, based on query interpretation and document annotations in the corpus. Each component is discussed in detail in the following subsections. 

\subsection{Query Processor}\label{qinterpret}
\begin{figure}[htp]
	\centering
		\caption{Workflow for translation of user queries into system queries}\label{fig:qpwork}
		\includegraphics[scale=0.50]{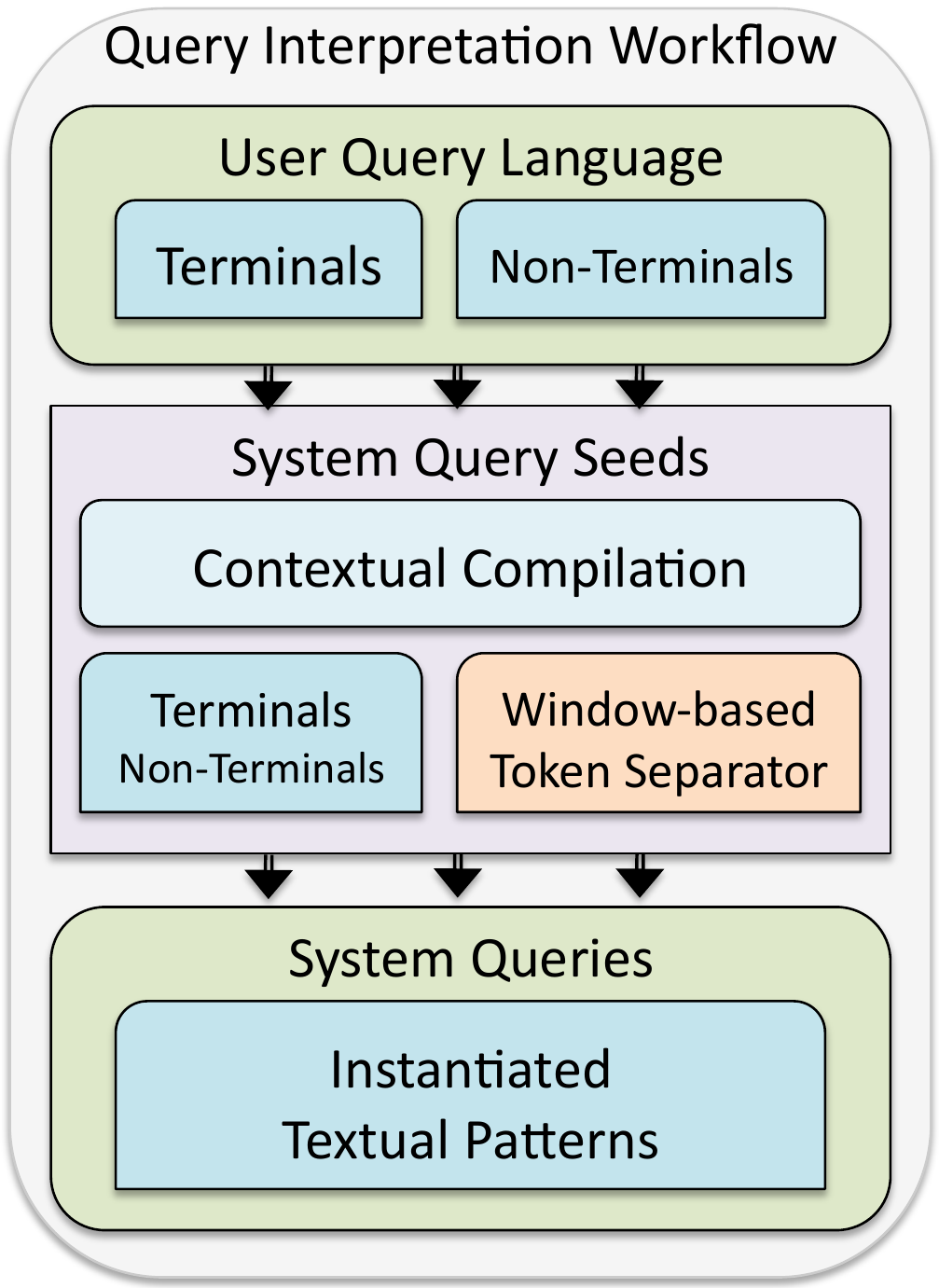}
\end{figure}

The process of searching for information from text commonly involves certain interactions between a user and a system. Users typically possess a conceptualization of their information need that can be framed using a mental model, as noted by Tran et al. \cite{TranCRS07}. The search system must provide an environment for users to adequately express their information need in terms of language primitives or a \textit{user query language} that can be understood by the system (Figure \ref{fig:qpwork}, top left). The system must then provide a specification for translating the user query into a \textit{system query} (Figure \ref{fig:qpwork}, left center), based on the interpretation of the user query. In our application, the query processor provides a \textit{user query interface} for users to specify their queries. It then performs the translation from user query into system query based on the underlying specification in the grammar. To account for domain specific constructs in the knowledge model, user queries are specified using templates, instead of free-form queries. These templates abstract data from the aforementioned (four) categories of data elements, which are important to the domain. The CFG is presented in the next section.

\subsubsection{Context-Free Grammar}\label{grammardef}
The context-free grammar used in our hybrid information retrieval system is formally defined in this section, along with anecdotal examples to illustrate how it is used in practice. Definitions 1-3 cover the top-level grammar, while Definitions 4-7 cover the low-level grammar. 

\newtheorem{mydef}{Definition}
\begin{mydef}\label{def1}
The context-free grammar G for the query language U of the hybrid information retrieval system H is a quadruple (N, T, P, S), where N is a finite set of non-terminals, T is a finite set of terminals (or alphabet), P is a finite set of rules (or productions) and S is a Start Symbol. 
\end{mydef}

The set of nonterminals $N$ is partitioned into two sets, $N^S$ and $N^P$, where $N^S$ denotes the set of nonterminals found directly in the right-hand sides (RHS) of the productions associated with the start symbol $S$, and $N^P = N - N^S$, where the symbol `--' is the set-difference operator. The set  $N^S$ contains 11 nonterminals, including {\footnotesize $\langle$DOSAGE$\rangle$, $\langle$FREQUENCY$\rangle$} and route of administration {\footnotesize $\langle$ROA$\rangle$} (see Table \ref{tab:templateclasses}, Section \ref{knomo} and \ref{grammarappendix}), which abstracts broad template classes of data relevant to the domain. The user query language of $H$ is formally defined as follows:  

\begin{mydef}\label{def2}
The user query language U of the hybrid information retrieval system H is the set of sentential forms over ($N^S \cup T )^\ast$ derivable from G = (N, T, P, S). That is, a user query q, may consist of terminals and nonterminals that appear only in the start symbol productions.
\end{mydef}

\begin{table*}
	\caption{Derivation of system query strings using the CFG }
	\centering
 	\label{tab:queryderiv}
	    \begin{tabular}{ | l | l  | l | l | l | l | c | c | l | c |} \hline
			1 & {\footnotesize $\langle$TEMPLATE PATTERN$\rangle$} & \multicolumn{2}{l|}{\footnotesize {\footnotesize ::= $\langle$ENTITY$\rangle$}} & \multicolumn{2}{l|}{{\footnotesize $\langle$PRONOUN$\rangle$}} & \multicolumn{2}{c|}{{\footnotesize $\langle$DOSAGE$\rangle$}}  & {\footnotesize $\langle$INTENSITY$\rangle$} \\ \hline
			2 & User Query &  \multicolumn{2}{l|}{{\footnotesize ::= $\langle$Buprenorphine$\rangle$}} & \multicolumn{2}{l|}{{\footnotesize $\langle$PERSONAL PRONOUN$\rangle$}} & \multicolumn{2}{c|}{\footnotesize ``$>$4mg''}  &  {\footnotesize $\langle$BY DAY $|$ BY HOUR$\rangle$} \\\hline
			3 & System Query Seed & {\footnotesize ::= $\langle$Subs$\rangle$}  &  {\footnotesize $\langle$RANGE$\rangle$}  &  {\footnotesize I} &  {\footnotesize $\langle$RANGE$\rangle$} &  {\footnotesize ``32mg''} &  {\footnotesize $\langle$RANGE$\rangle$}  &  {\footnotesize``a day''} \\\hline
			4 & System Query & \multicolumn{7}{c|}{{\footnotesize ::= Subs I \underline{was taking} 32mg a day}}\\\hline
			\multicolumn{9}{|c|}{}  \\\hline

			5 & {\footnotesize $\langle$TEMPLATE PATTERN$\rangle$} & \multicolumn{2}{l|}{\footnotesize {\footnotesize ::= $\langle$ENTITY$\rangle$}} & \multicolumn{2}{l|}{{\footnotesize $\langle$PRONOUN$\rangle$}} & \multicolumn{2}{c|}{{\footnotesize $\langle$DOSAGE$\rangle$}}  & {\footnotesize $\langle$INTENSITY$\rangle$} \\ \hline
			6 & User Query &  \multicolumn{2}{l|}{{\footnotesize ::= $\langle$vicodin$\rangle$}} & \multicolumn{2}{l|}{{\footnotesize $\langle$PERSONAL PRONOUN$\rangle$}} & \multicolumn{2}{c|}{\footnotesize ``$>$28mg''}  &  {\footnotesize $\langle$BY DAY $|$ BY HOUR$\rangle$}\\\hline
			7 & System Query Seed & {\footnotesize ::= $\langle$vicodin$\rangle$}  &  {\footnotesize $\langle$RANGE$\rangle$}  &  {\footnotesize I} &  {\footnotesize $\langle$RANGE$\rangle$} &  {\footnotesize ``32mg''} &  {\footnotesize $\langle$RANGE$\rangle$}  &  {\footnotesize ``every day''} \\\hline
			8 & System Query & \multicolumn{7}{c|}{{\footnotesize ::= vicodin \underline{habit and} I \underline{was taking} 28mg \underline{of buprenorphine} every day}} \\\hline
	    \end{tabular}
\end{table*}

\noindent For example, the production: {\footnotesize $\langle$TEMPLATE PATTERN$\rangle$ $\rightarrow$ $\langle$ENTITY$\rangle$  $\langle$PRONOUN$\rangle$ $\langle$DOSAGE$\rangle$  $\langle$INTENSITY$\rangle$}  is a broad template pattern that abstracts the information need given in Section \ref{intro} in $G$, where {\footnotesize $\langle$TEMPLATE PATTERN$\rangle$} is the start symbol, {\footnotesize $\langle$ENTITY$\rangle$,  $\langle$PRONOUN$\rangle$, $\langle$DOSAGE$\rangle$} and {\footnotesize $\langle$INTENSITY$\rangle$} are nonterminals, or template classes, in $N^S$. The more specific user query: {\footnotesize $\langle$q$\rangle$ ::= $\langle$Buprenorphine$\rangle$  $\langle$PERSONAL PRONOUN$\rangle$ ``$>$4mg''  $\langle$BY DAY $|$ BY HOUR$\rangle$} is a valid user query for the given information need, derived from this production, where {\footnotesize $\langle$Buprenorphine$\rangle$} is a member of the template class {\footnotesize $\langle$ENTITY$\rangle$}, {\footnotesize $\langle$PERSONAL PRONOUN$\rangle$} is a member of the template class {\footnotesize $\langle$PRONOUN$\rangle$}, {\footnotesize $\langle$BY DAY$\rangle$}  and {\footnotesize $\langle$BY HOUR$\rangle$} are non-terminals, derived from the template class {\footnotesize $\langle$INTERVAL$\rangle$}. The expression ``$>$4mg'' is a terminal, which requires special interpretation (discussed later in this section). User queries may therefore consist of permutations of terminals and nonterminals, or sentential forms in $G$. Lines 2 and 6 in Table \ref{tab:queryderiv}, show two valid user  queries derived from the given production. Lines 4 and 8 show two specific system queries (which are matching text snippets in the corpus) derived from the user queries in lines 2 and 6. 

The translation of user queries into system queries is a two-step process. \textit{System query seeds} must first be generated from user queries (Figure \ref{fig:qpwork}, center) and then  transformed into system queries by instantiating a window-based token separator. For instance, the system query seed:  ``Subs {\footnotesize $\langle$RANGE$\rangle$} I {\footnotesize $\langle$RANGE$\rangle$} 32mg {\footnotesize $\langle$RANGE$\rangle$} a day,'' is generated from the user query:  {\footnotesize $\langle$q$\rangle$ $\rightarrow$ $\langle$Buprenorphine$\rangle$  $\langle$PERSONAL PRONOUN$\rangle$ ``$>$4mg''  $\langle$BY DAY $|$ BY HOUR$\rangle$}, where ``Subs'' is a synonym for ``Buprenorphine,'' ``I'' is a personal pronoun, ``32mg'' is greater than ``4mg'' and ``a day'' is an expression for ``by day''. Upon instantiating the three successive {\footnotesize $\langle$RANGE$\rangle$} values that capture window size to 0, 2 and 0 respectively, the specific system query: ``Subs I \underline{was taking} 32mg a day,'' can be obtained. The sequence of tokens that occupy the range are shown in underline. Given this example, the interpretation of a user query is therefore defined as follows: 

\begin{mydef}\label{def3}
The interpretation of a user query q $\in$ U $\subseteq$ ($N^S \cup T )^\ast$  is a set of all terminal strings derivable from q in the grammar Q = (N, T, P, q), where the Start Symbol is replaced by q, a single sentential form obtained over ($N^S \cup T )$.
\end{mydef}

The previous definitions cover the top-level grammar. However, the precise translation of a user query into system query seeds requires additional preliminaries, especially to interpret compound expressions such as ``$>$4mg.'' The {\footnotesize DOSAGE-OPERATOR}, {\footnotesize DOSAGE-AMOUNT} and {\footnotesize DOSAGE-UNIT} have instantiations that require appropriate expansion. For example, ``much more than 4mg'', ``five mg,'' ``60 milligrams'' and ``a hundred milligrams'' are valid  interpretations for the query fragment ``$>$4mg.'' To capture this, we introduce the notion of a \textit{contextual compilation}  (Figure \ref{fig:qpwork}, center) to formalize the translation of any terminal to its semantic equivalent according to its interpretation in $G$. Let  \textit{cc(``$>$4mg,'' {\footnotesize $\langle$DOSAGE$\rangle$})}  denote the contextual compilation of the expression ``$>$4mg,'' which belongs to the class {\footnotesize $\langle$DOSAGE$\rangle$}. Formally: 

\begin{mydef}\label{def4}
A contextual compilation cc of a terminal string t derived from a nonterminal A is the set of terminal strings semantically equivalent to t in the context of A.
\end{mydef}

\noindent It follows then that the interpretation of a user query therefore requires interpretation of both nonterminals and terminals, whenever the latter contains equivalent interpretations. 

\begin{mydef}\label{def5}
The translation $\Gamma(t)$ of a terminal t derivable from the nonterminal $A \in N^S$ in G = (N, T, P, S) is its contextual compilation $\Gamma(t) = cc(t, A)$.
\end{mydef}

\begin{mydef}\label{def6}
The translation $\Gamma(A)$ of a nonterminal A in the grammar G = (N, T, P, S) is defined as the set of terminal strings that can be derived from A. That is,  $\Gamma(A) = \lbrace t$  $|$ $A \Rightarrow_{\tiny G}^{\ast} t\rbrace$
\end{mydef}

\noindent Note that this translation of terminals can be specified by domain experts or by search engine developers programmatically.  For example, the translation of the `greater than' operator is specified explicitly as: $\langle greaterThanOp\rangle$ $\rightarrow$ \textbf{$>$ $|$ greater than $|$ more than $|$ above $|$ in excess of $|$ \ldots} in the grammar. Given the definition of a user query, system query seeds can then be formally defined based on the translation of user query elements and the window-based {\footnotesize $\langle$RANGE$\rangle$} token separator.

\begin{mydef}\label{def7}
The system query seeds of a user query $q =\alpha_1, \alpha_2, \ldots, \alpha_n$ where $\alpha_i \in (N^S \cup T)$, is the cross-product of a the translation of the terminals and nonterminals $\Gamma(\alpha_1)$ $\times$ $\Gamma(\alpha_2)$ $\times$ $\ldots$ $\times$ $\Gamma(\alpha_n)$ that comprise the user query. 
\end{mydef}

As noted, system query seeds become actual system queries when the {\footnotesize $\langle$RANGE$\rangle$} operator is instantiated. For example, Table \ref{tab:queryderiv} shows how the two system queries: 1) ``Subs I was taking 32mg a day'' and ``vicodin habit and I take 28mg of buprenorphine every day'' could be derived from the production:  {\footnotesize $\langle$TEMPLATE PATTERN$\rangle$ $\rightarrow$ $\langle$ENTITY$\rangle$  $\langle$PRONOUN$\rangle$ $\langle$DOSAGE$\rangle$  $\langle$INTENSITY$\rangle$}, based on the grammar. The production contains nonterminals in the RHS, which are in $N^S$ called \textit{template classes}. The user first selects the broad template pattern and then constructs a more specific user query, where appropriate. The grammar then generates the system query seeds, which become system queries after instantiating the {\footnotesize $\langle$RANGE$\rangle$} values. Documents that contain textual patterns that match system queries in the corpus are considered as candidate matches for the user query.

Note that the system query for the second user query (Table \ref{tab:queryderiv}, line 8) could also be considered a match for first user query in Table \ref{tab:queryderiv}, line 2. This is because initially, all annotations are retrieved for a given template and then filtered by the query matcher. The combination of ``vicodin'' as an {\footnotesize $\langle$ENTITY$\rangle$} in the first position, separated by a  {\footnotesize $\langle$PERSONAL PRONOUN$\rangle$} (``I''), then a {\footnotesize $\langle$DOSAGE$\rangle$} exceeding the prescribed limit (``28mg'') and then an {\footnotesize $\langle$INTENSITY$\rangle$} (``every day''), could be a match because ``buprenorphine'' appears as a concept in one of the {\footnotesize $\langle$RANGE$\rangle$} separators. In the next section we discuss how the four categories of data are represented in the knowledge model and interpreted by the grammar. 

\subsubsection{Knowledge Model}\label{knomo}
In Definition 1, we described the set of nonterminals $N^S$ as the set of nonterminals in the RHS of productions that begin with the start symbol {\footnotesize $\langle$TEMPLATE PATTERN$\rangle$}. For example, in the production: {\footnotesize $\langle$TEMPLATE PATTERN$\rangle$ $\rightarrow$ $\langle$ENTITY$\rangle$  $\langle$PRONOUN$\rangle$ $\langle$DOSAGE$\rangle$  $\langle$INTENSITY$\rangle$}, the RHS elements {\footnotesize $\langle$ENTITY$\rangle$,  $\langle$PRONOUN$\rangle$, $\langle$DOSAGE$\rangle$} and {\footnotesize $\langle$INTENSITY$\rangle$} are in the set $N^S$. According to the grammar, $N^S$ consists of 11 nonterminals or template classes (shown in Table \ref{tab:templateclasses}). These classes cover the four categories of data interpretable by the system: 1) concepts in ontologies; 2) concepts in lexicons; 3) lexico-ontology concepts and 4) intelligible constructs specified using rules. The ability to interpret these categories of data is a key contribution, which is crucial to the effectiveness of our system for domain specific information retrieval. In the next section, we begin with the interpretation of the template class, which are based on the ontology. 

\begin{table}
	\caption{Template Class Classification}
	\centering
 	\label{tab:templateclasses}
	    \begin{tabular}{ | c | l  | c | c |} \hline
			\multicolumn{2}{|l|}{\multirow{1}{*}{\textbf{Template Class Name}}} & \multirow{1}{*}{\textbf{Class Source}} & \multirow{1}{*}{\textbf{Class Type}} \\ \hline
			1& {\footnotesize $\langle$INTERVAL$\rangle$} & Alphabet & Compound \\ \hline
			2 & {\footnotesize $\langle$DOSAGE$\rangle$} & Alphabet  & Compound \\ \hline
			3 & {\footnotesize $\langle$FREQUENCY$\rangle$} &  Alphabet  & Compound \\ \hline
			4 & {\footnotesize $\langle$ENTITY$\rangle$} & Ontology &  Simple \\ \hline
			5 & {\footnotesize $\langle$ROA$\rangle$} & Lexico-ontology & Simple \\ \hline
			6 & {\footnotesize $\langle$DRUGFORM$\rangle$} & Lexico-ontology & Simple \\ \hline
			7 & {\footnotesize $\langle$SIDEEFFECT$\rangle$} & Lexico-ontology & Simple \\ \hline
			8 & {\footnotesize $\langle$EMOTION$\rangle$} & Lexicon & Simple \\ \hline
			9 & {\footnotesize $\langle$PRONOUN$\rangle$} & Lexicon & Simple \\ \hline
			10 & {\footnotesize $\langle$INTENSITY$\rangle$} & Lexicon & Simple \\ \hline
			11 & {\footnotesize $\langle$SENTIMENT$\rangle$} & Lexicon & Simple \\ \hline
	    \end{tabular}
\end{table}

\textbf{Ontology-based Query Interpretation}: To facilitate ontology-based query interpretation, we utilize a Drug Abuse Ontology (DAO) (\textit{pronounced dow}) \cite{CameronSDSDCACWF13}, which was created as part of the PREDOSE project. The DAO consists of 43 classes and 20 properties, and serves two main purposes. First, it facilitates query interpretation, and second it serves as one of the annotation schemes for metadata extraction (discussed in Section \ref{docannot}).  The DAO is important for query interpretation because it captures various mappings between slang terms and standard drug references. In a gold standard dataset consisting of 600 web forum posts, we observed a ratio of 33:1 slang references for the standard drug label for the prescription drug ``Buprenorphine'' and 24:1  for ``Loperamide.''  The DAO is therefore of critical importance. 

To perform such query interpretation based on the DAO, let the drug abuse ontology $O$ be represented as a graph $O = (V, E)$, where $V$ is the set of nodes, which formally represent real-world concepts $V =  \lbrace v_1, v_2, \ldots, v_n\rbrace$ and $E = \lbrace e_1, e_2, \ldots, e_m \rbrace$ is the set of edges, which represent labeled edges (or relationships) between such concepts. The interpretation of a keyword $k_i$ in $q$ according to the ontology O, denoted $I(k_i)$, is some concept $v_i$ in $O$, where the concept $v_i$ may be a class $c_i$ or an instance $r_i$. That is, $v_i \in (C \cup R)$, such that $c_i \in C$ and $r_i \in R$, where $C$ is the set of all ontology classes and $R$ is the set of all instances. The label of the class $c_i$ is denoted $L(c_i)$ and the label of an instance is denoted $L(r_i)$. The set of all labels for all classes in $O$ is denoted $L(C)$, while set of all slang terms for all classes is $L_s(C)$. Similarly the set of all labels for instances is $L(R)$ and their slang terms $L_S(R)$, respectively. 

A keyword $k_i$ in the corpus maps to a class or an instance level concept in the DAO. This interpretation is based on string matching using the labels and synonyms  of instances and classes. String matching is sufficient, since the overlap in concept labels in the DAO is relatively small. Evidence for this comes from the evaluation of our entity identification approach in \cite{CameronSDSDCACWF13}, in which 85\% of slang term mentions in the gold standard could be easily reconciled to the correct ontology concept, without disambiguation. In the evaluation (see Section \ref{eval}), we show explicitly that it is the ability to interpret intelligible constructs not captured by ontologies that is more crucial to domain specific information retrieval and concepts from the DAO are abstracted using the template class {\footnotesize $\langle$ENTITY$\rangle$} (shown in Table \ref{tab:templateclasses}, row 4).

\textbf{Lexico-ontology-based Query Interpretation:} Lexico-ontology concepts are those that have partial representation in ontologies and lexicons simultaneously. For example,  the side effects ``skin blisters that are itchy'' and ``skin blisters that are painful'' are distinct ``skin blisters.'' However, an ontology may only contain the side effect ``skin blisters.'' This may be problematic if the distinction between itches and pain contribute new information about trends in drug abuse. Knowledge of mappings from lexicons, not present in the ontology, could therefore improve the effectiveness of the search system. 

In practice, such discrepancies between lexicons and ontologies for the same concept, may arise as a natural consequence of ontology evolution. Various concepts (or additional attributes for existing ones), may eventually be added to the ontology, but should not be excluded from the search framework in the interim. Also, references in lexicons (such as the Urban Disctionary) may be unknown to domain experts altogether, and never come under consideration for formal representation in the ontology. Inclusion of such search terms in the system may bridge this knowledge gap between what is modeled and what is evident within the community. To address this, we introduce a category of concepts called \textit{lexico-ontology} concepts. The three template classes: route of administration {\footnotesize $\langle$ROA$\rangle$}, {\footnotesize $\langle$DRUGFORM$\rangle$} and {\footnotesize $\langle$SIDEEFFECT$\rangle$} are lexico-ontology concepts in our system (Table \ref{tab:templateclasses}, rows 5-7). 

\textbf{Lexicon-based Query Interpretation:} The four template classes:   {\footnotesize $\langle$SENTIMENT$\rangle$},  {\footnotesize $\langle$EMOTION$\rangle$},  {\footnotesize $\langle$PRONOUN$\rangle$} and  {\footnotesize $\langle$INTENSITY$\rangle$} (Table \ref{tab:templateclasses}, rows 8-11) are part of an ubiquitous class of nonterminals present in many lexicons. These classes provide insights into self-disclosures, opinions, reports on mood changes and various other experiences in response to drug use. Sentiment expressions such as ``didnt do sh*t,'' ``not that great'' and ``felt pretty good'' can help epidemiologists assess and evaluate user reaction and attitudes. In our application, {\footnotesize $\langle$SENTIMENT$\rangle$} expressions are classified into three categories: {\footnotesize $\langle$POSITIVE$\rangle$}, {\footnotesize $\langle$NEGATIVE$\rangle$} and {\footnotesize $\langle$NEUTRAL$\rangle$} based on several lexicons, including LIWC\footnote{LIWC Online -- http://www.liwc.net/} and MPQA\footnote{MPQA -- http://mpqa.cs.pitt.edu/}. The sentiment identification algorithm implemented by Chen et. al. in \cite{ChenWNWS12} is then used to link and annotate sentiment expressions in the corpus. To interpret {\footnotesize $\langle$EMOTION$\rangle$}, the online resource \textit{ChangingMinds.org} that contains several categorizes, such as {\footnotesize $\langle$AFFECTION$\rangle$}, {\footnotesize $\langle$LUST$\rangle$}, {\footnotesize $\langle$LOVE$\rangle$} and {\footnotesize $\langle$RAGE$\rangle$} is used. Similarly, various online lexicons are used for categorization of such classes of {\footnotesize $\langle$PRONOUN$\rangle$} including {\footnotesize $\langle$PERSONAL PRONOUN$\rangle$}, {\footnotesize $\langle$INTERROGATIVE PRONOUN$\rangle$} and {\footnotesize $\langle$POSSESSIVE PRONOUN$\rangle$}. Distinguishing self references, which are of type {\footnotesize $\langle$PERSONAL PRONOUN$\rangle$}, are important in identifying drug users as the subject of discussions. The interpretation of the nonterminal {\footnotesize $\langle$INTENSITY$\rangle$} is based on the domain. Expressions such as ``low,'' ``small'' and ``less'' can convey {\footnotesize $\langle$LOW$\rangle$} intensity with regards to drug usage, while ``largest,'' ``excessive'' and ``most'' can convey {\footnotesize $\langle$HIGH$\rangle$} intensity (see \ref{grammarappendix}).  

\textbf{Rule-based Query Interpretation:} The three nonterminals {\footnotesize $\langle$INTERVAL$\rangle$}, {\footnotesize $\langle$FREQUENCY$\rangle$} and {\footnotesize $\langle$DOSAGE$\rangle$} (Table \ref{tab:templateclasses}, rows 1-3) require more complex interpretation, and are considered compound template classes. For example, the derivation of a system query from the class {\footnotesize $\langle$INTERVAL$\rangle$} can be constructed as follows: {\footnotesize $\langle$INTERVAL$\rangle$} $\rightarrow$ {\footnotesize $\langle$AMOUNT$\rangle$} {\footnotesize $\langle$DURATION\_INDICATOR$\rangle$} {\footnotesize $\langle$PERIOD\_DETERMINER$\rangle$}. In this production, a valid {\footnotesize $\langle$INTERVAL$\rangle$} consists of any {\footnotesize $\langle$AMOUNT$\rangle$} (numeric or textual), followed by a {\footnotesize $\langle$DURATION\_INDICATOR$\rangle$} (e.g. ``days,'' ``weeks,'' ``years'') and a {\footnotesize $\langle$PERIOD\_DETERMINER$\rangle$} (e.g. ``now,'' ``before,'' ``ago''). The system queries ``5 years ago'' and ``about nine months later'' are therefore valid interval expressions. Similarly, the following production: {\footnotesize $\langle$FREQUENCY$\rangle$} $\rightarrow$ {\footnotesize $\langle$AMOUNT$\rangle$} {\footnotesize $\langle$PER\_TIME\_INDICATOR$\rangle$}, can be used to derive valid {\footnotesize $\langle$FREQUENCY$\rangle$} expressions, such as ``5 per min,'' ``per hour'' and ``24 mg /min.''  {\footnotesize $\langle$DOSAGE$\rangle$} system queries such as ``1-5 grams'' and ``2 mcg'' can be derived according to the grammar, based on the following production: {\footnotesize $\langle$DOSAGE$\rangle$} $\rightarrow$ {\footnotesize $\langle$NUMBER\_AMOUNT$\rangle$}  {\footnotesize $\langle$UNIT$\rangle$} (see \ref{grammarappendix}) for a partial list of productions in the grammar. 

The grammar consists of of 61 template patterns in the top-level CFG\footnote{Top-Level Grammar Productions \url{http://wiki.knoesis.org/index.php/Knowledge-Aware-Search-Productions}}, consisting of template classes in $N^S$ in the RHS of the productions. It also contains close to 150 productions in $N^P$ (see partial list in \ref{grammarappendix}). Using this grammar, our search system in PREDOSE is able to perform domain specific information retrieval somewhat effectively compared with existing search systems (see Section \ref{eval}). The top-level grammar enables query specification according to the direct information needs of epidemiologists, while the low-level grammar enables interpretation of four different categories of data, pertinent to the domain.  In the next section we discuss the use of this grammar for metadata extraction and document annotation from the corpus. 

\subsection{Semantic Metadata Extractor}\label{docannot}
The semantic metadata extractor (Figure \ref{fig:architecture}, right) identifies textual patterns (i.e., system queries) in the corpus that match the productions in the grammar. The extractor maintains a database of mappings between these textual patterns in the corpus and web forum posts, which contain them.  All annotation extraction is performed offline in a pre-processing step. Given a system query, the \textit{query matcher} (Figure \ref{fig:architecture}, bottom left) retrieves the matching documents after applying various filters (discussed in Section \ref{qmatch}).

\begin{figure}[htp]
	\centering
		\caption{Sample AQL queries}\label{fig:aql}
		\includegraphics[scale=0.70]{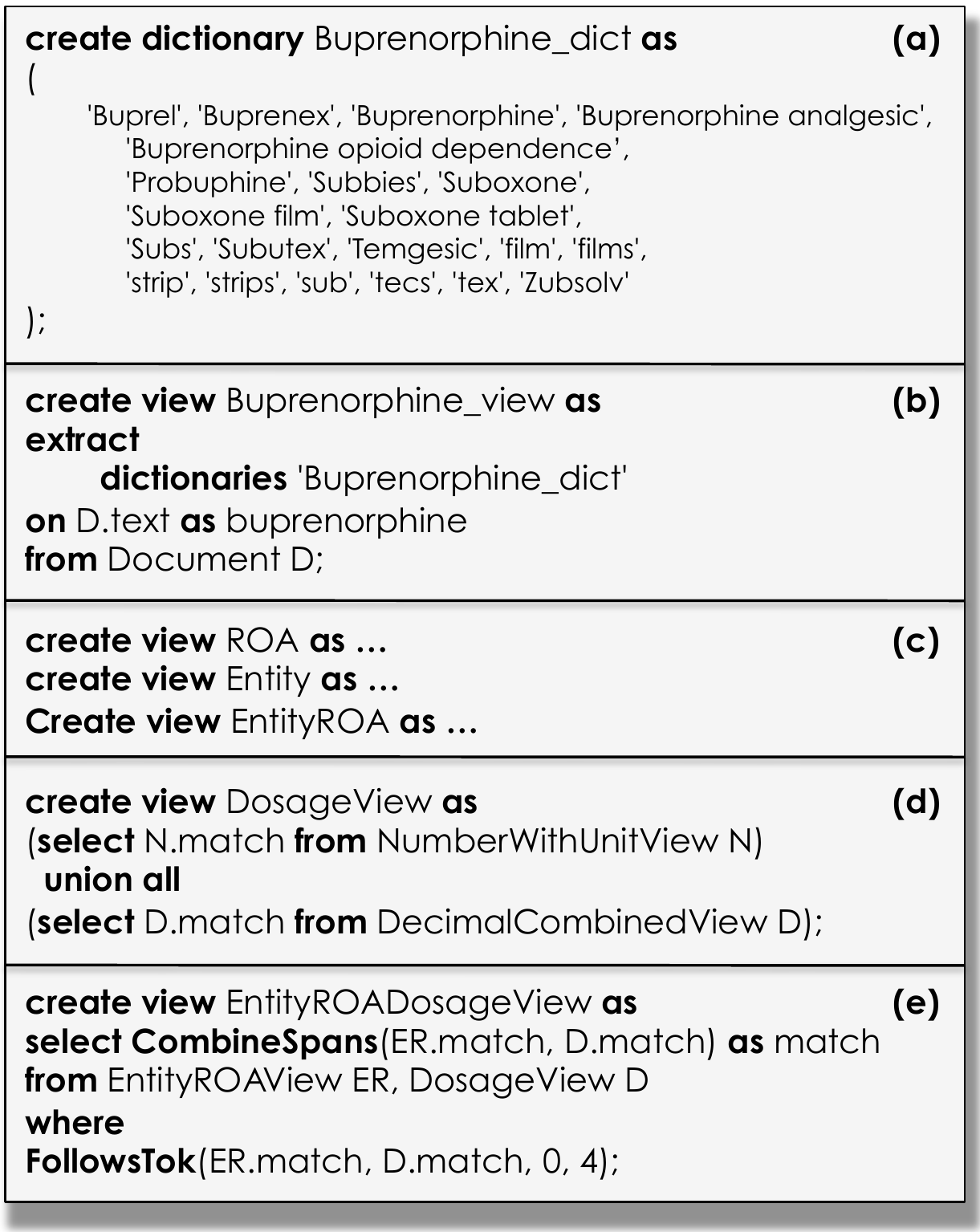}
\end{figure}

To retrieve the semantic metadata from the corpus we rely on the SystemT \cite{Chiticariu2010, KrishnamurthyLRRVZ08} framework, and its declarative language specification -- AQL (Annotation Query Language). SystemT is a scalable algebraic framework for extracting structured information from unstructured text. It abstracts and manipulates textual data using relational operators such as \textit{select, join, union and consolidate}. Queries can be formulated using AQL then executed with SystemT. AQL is based primarily on two constructs: the \textit{view} and the \textit{dictionary}. An AQL dictionary contains a list of strings and can be exposed as a view. For example, the `Buprenorphine' dictionary (\textit{Buprenorphine\_dict}) shown in Figure \ref{fig:aql}(a) contains several synonyms for this concept obtained from the DAO. This dictionary can then be exposed as the \textit{Buprenorphine\_view} as shown in Figure \ref{fig:aql}(b). More complex views (or patterns) can be derived by nesting existing views. Figure \ref{fig:aql}(e) shows the AQL query that extracts textual patterns for the production:  {\footnotesize $\langle$TEMPLATE PATTERN$\rangle$ $\rightarrow$ $\langle$ENTITY$\rangle$  $\langle$ROA$\rangle$ $\langle$DOSAGE$\rangle$}, where {\footnotesize $\langle$ROA$\rangle$} refers to route of administration (ROA). The \textit{EntityROADosageView} is an annotator in which an ``{\footnotesize $\langle$ENTITY$\rangle$} {\footnotesize $\langle$ROA$\rangle$}'' pattern must occur within 4 tokens of a {\footnotesize $\langle$DOSAGE$\rangle$} expression. As shown in Figure \ref{fig:aql}(d), a {\footnotesize $\langle$DOSAGE$\rangle$} expression can be defined as any numerical value that co-occurs with a unit, or any decimal value combined with a unit (see \ref{grammarappendix}). 

All AQL queries are written, compiled and published for execution on the corpus using the IBM BigInsights platform. Then the Unstructured Information Management Architecture (UIMA) \cite{uima} is used to execute the queries on the corpus. The annotated corpus contained 1,287,830 annotations from a corpus of approximately 1,026,502 web forum posts from three online forums\footnote{Please note that in compliance with the Institutional Review Board (IRB) protocol approved for the PREDOSE project at the Wright State University, to which we are required to adhere to, the names of the selected sources have not been disclosed.}. The extracted metadata was indexed positionally (using Apache Lucene and Solr) for use by the query matcher, and also to provide highlighted annotations in the search results. Techniques for matching queries with documents based on extracted semantic metadata are discussed in the following section. 

\subsection{Query Matcher}\label{qmatch}
The query matcher (Figure  \ref{fig:architecture}, bottom left) retrieves relevant documents based on a match between a system query and an annotation extracted using the semantic metadata extractor. To achieve this, the system adopts a two-step process. First, the query processor selects all documents indexed with the template pattern of the user query. For example, given the user query: {\footnotesize $\langle$q$\rangle$} $\rightarrow$ {\footnotesize $\langle$Buprenorphine$\rangle$} {\footnotesize $\langle$PERSONAL PRONOUN$\rangle$} {\footnotesize ``$>$4mg''} {\footnotesize $\langle$BY DAY$\rangle$} $|$ {\footnotesize $\langle$BY HOUR$\rangle$}, derived from template pattern: {\footnotesize $\langle$TEMPLATE PATTERN$\rangle$} $\rightarrow$ {\footnotesize $\langle$ENTITY$\rangle$} {\footnotesize $\langle$PRONOUN$\rangle$} {\footnotesize $\langle$DOSAGE$\rangle$} {\footnotesize $\langle$INTENSITY$\rangle$}, the query processor selects an initial set of documents (518) that contain the matching annotations for the given template pattern. Second, the query matcher applies various filters to prune the search results. The \textit{EntityFilter} is first used to retain documents containing {\footnotesize $\langle$Buprenorphine$\rangle$}, specified by ontology-driven query interpretation in the grammar. The resulting document set is reduced (to 97). The query matcher then applies the  \textit{PronounFilter}, which restricts the result set to annotations containing only {\footnotesize $\langle$PERSONAL PRONOUN$\rangle$} (resulting in 90 documents). The query matcher then applies the  \textit{DosageFilter}, which retains annotations that mention amounts greater than ``4mg,'' according to the interpretation in the grammar. Recall that this translation involves interpreting of the contextual compilation of the expression ``$>$4mg,'' which requires interpretation of the greater than ``$>$'' operator based on synonyms (e.g., ``greater than'' and ``more than''), mapping the numeral ``4'' to the word ``four,'' and also expanding the unit ``mg'' with its various semantically equivalent forms (``milligram,'' ``mgs'' and ``milli-grams''). The resulting document set is then reduced (to 40). Finally, the query matcher applies the  \textit{IntervalFilter}, which restricts the document set to only those annotations that mention daily use {\footnotesize $\langle$BY DAY$\rangle$} $|$ {\footnotesize $\langle$BY HOUR$\rangle$}. The result is a document set consisting of 21 documents. The ability to extract such search results, based on the grammar, is key to effective domain specific search. In the next section we discuss the user-driven evaluation of our hybrid search system based on the application of our overall search paradigm to the domain of prescription drug abuse. 

\section{Evaluation}\label{eval}
Search systems are typically evaluated using  precision, recall and F-Score metrics computed against a baseline of relevant document, for various queries. However, in prescription drug abuse, gold standard datasets are unavailable. In general, the unavailability of standardized datasets for evaluating semantic search system is a common issue in the semantic web community \cite{Sure02, McCool05}. To evaluate our approach, we perform a comparative analysis of our system against existing search systems through a user-driven evaluation. We note that subjective differences in user agreement and relevance judgments may unduly impact the quality of the evaluation, as noted by Blanco et. al. \cite{BlancoHHMPTT13}. Still, the expectation is that our domain specific information retrieval system will perform better than existing search systems, for these domain specific searches. Hence, the goal of the evaluation is to assess the shortcomings of existing systems and stress the need for richer systems for domain specific searches. 

We selected three search systems for evaluation: 1) Hakia, 2) DuckDuckGo and 3) Google. Hakia was selected because it uses a SemanticRank algorithm  together with background knowledge for search, and therefore fits the characteristics of a classic semantic search engine. DuckDuckGo was selected because it is uses crowd-sourced data from Wikipedia, Wolfram Alpha and Bing (formerly Powerset). The popular search engine Google was selected due to its prominence in general purpose search. 

To conduct the evaluation we asked colleagues, not involved in this research but attached to the Kno.e.sis Center to participate in the user study\footnote{A live version of the search system is available online for option viewing -- http://knoesis-hpco.cs.wright.edu/knowledge-aware-search; please refer to the accompanying video demo for a system overview)}. Each query was executed on the same undisclosed web forum and provided \textit{a priori} to evaluators after a short tutorial of the system. Each evaluator was asked to evaluate the relevance of retrieved documents across all four systems. Initial relevance judgements were based on the text snippet in the search result. If deemed interesting, document should then be explored to confirm or disprove relevance. 

\begin{table}
	\caption{Evaluation: User Query Scenarios}
	\centering
 	\label{tab:scenarios}
	    \begin{tabular}{ | p{8.5cm} |} \hline
		 What specific information is being shared by individuals in the corpus on the use of Buprenorphine in dosages exceeding 4mg daily? \\ \hline
		 What negative sentiments (or experiences) are being conveyed in the corpus by individuals towards the use of Buprenorphine? \\ \hline
	    \end{tabular}
\end{table}
Two query scenarios were used in the evaluation (shown in Table \ref{tab:scenarios}). Each was then repeated once with different constraints. Thus, four scenarios were examined. These scenarios require interpretation of ontological concepts, concepts in lexicons, and rule-based derivations. Lexico-ontology concepts are not included in the evaluation, however we note that several queries for which they are relevant, exist in PREDOSE\footnote{Numerous search queries are available online for optional viewing -- http://wiki.knoesis.org/index.php/Knowledge-Aware-Search-Evaluation}.  

In the first query, which is ``What specific information is being shared by individuals in the corpus on the use of Buprenorphine in dosages exceeding 4mg daily?'', our system interprets the keyword ``buprenorphine'' using the ontology, the keyword ``daily'' using the lexicon, the keyword ``individuals'' using a lexicon, and the phrase ``dosages exceeding 4mg'' through rules in the grammar. In the second query, the keyword ``buprenorphine'' is again interpreted from the ontology, the keyword ``individuals'' is interpreted using a lexicon and the keyword phrase ``negative sentiments'' is interpreted using the sentiment lexicon (and the method by Chen et. al. \cite{ChenWNWS12}). The evaluators were asked to perform their evaluation by first dynamically formulating a query of their choice, for use in Google, Hakia and DuckDuckGo, but using a static query in PREDOSE. This dynamic query requirement was intended to capture the subjective viewpoints of the various evaluators. All measures are based on the top 20 hits in each search system. 
\begin{figure*}[ht]
	\centering
		\caption{Evaluation Scenarios: 1(top left), 2(top right), 3(bottom left), 4(bottom right)}\label{fig:eval}
		\includegraphics[scale=0.65]{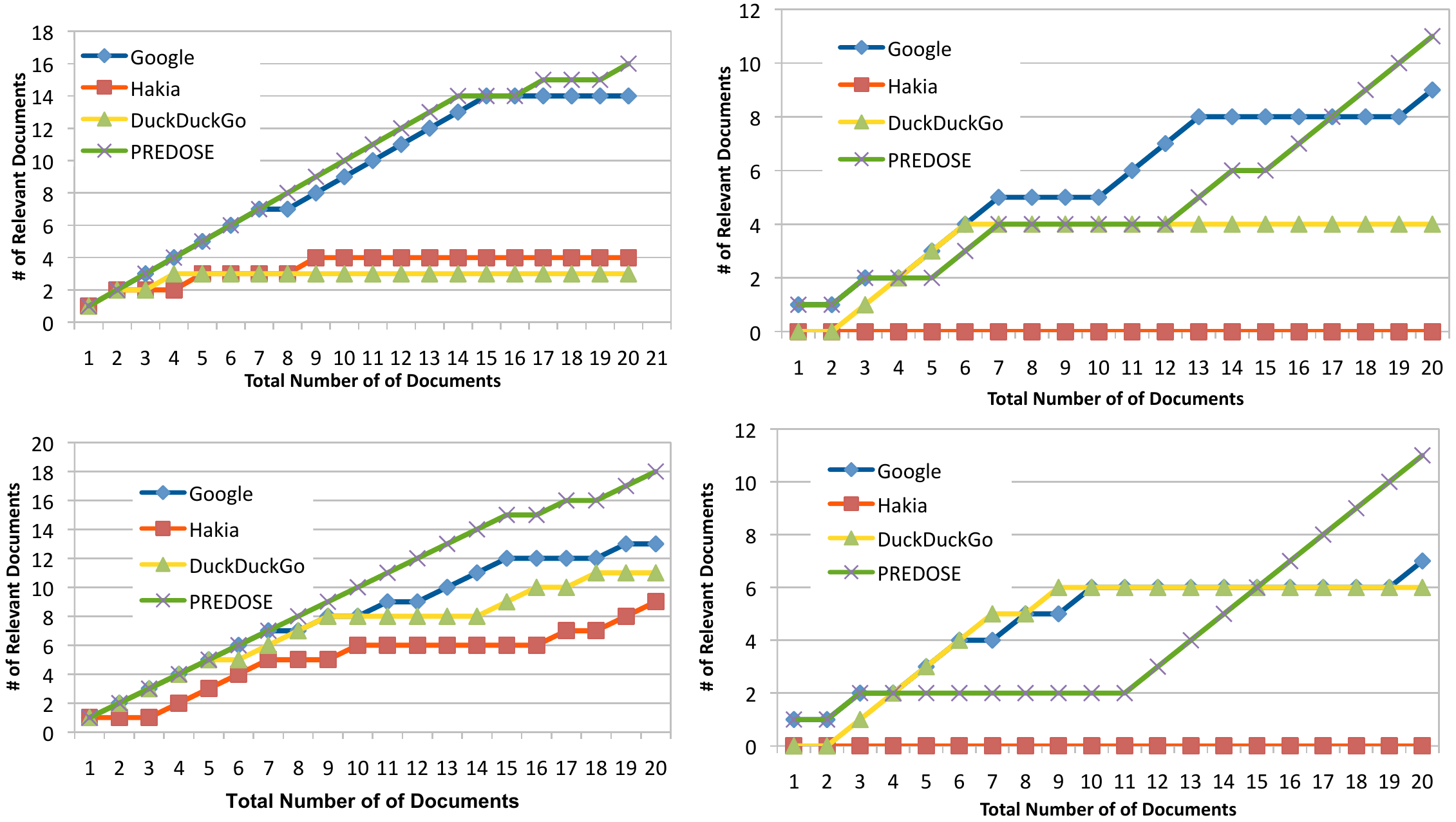}
\end{figure*}

\textbf{Scenario 1:} Six evaluators completed the evaluation by formulating an appropriate query of their choice for the web searches (i.e., Google, Hakia and DuckDuckGo), but using the following specific user query to search PREDOSE: {\footnotesize $\langle$Buprenorphine$\rangle$} {\footnotesize $\langle$PERSONAL PRONOUN$\rangle$} {\footnotesize ``$>$4mg''} {\footnotesize $\langle$BY DAY$\rangle$} $|$ {\footnotesize $\langle$BY HOUR$\rangle$}. Table \ref{tab:uqs1} shows that among the redacted  search queries for each user, each contained a mention of the keyword ÔbuprenorphineÕ and various expressions for excessive dosage, including the greater than operator ``$>$,'' ``more than,'' ``dosage excess,'' and ``over.'' 

Figure \ref{fig:eval} (top left) shows the results across the four systems. Our system retrieved 16/20 relevant results across the six evaluators. Google performed second best, by retrieving 14/20 relevant (but different) documents according to the human judgments. The Google search results showed that it was indeed able to retrieve documents with semantic equivalents for `buprenorphine' (namely `Suboxone' and `bupe'). However, the variability in our system was much greater. In particular, Google did not highlight any search result which contained a dosage greater than Ô4mg.Õ Instead, greater amounts occurred serendipitously in the snippet of search results. This is in stark contrast to our system in which all documents met this constraint. Furthermore, the result set in our system contained a few mentions of dosage in excess of  32mg, which is considered the known limit by epidemiologists. Hakia retrieved 4/20 relevant search results, which were not very informative. So were the search results from DuckDuckGo, which retrieved only 3/20 relevant results. On close inspection, we observed that their poor performance was largely due to an inability to interpret the semantics of the greater than (`$>$') operator. Most search results were retrieved because they contained the label `buprenorphine' itself and other query elements. 

\begin{table}
	\caption{User Queries for Scenario 1}
	\centering
 	\label{tab:uqs1}
	    \begin{tabular}{ | c | p{7.5cm} |} \hline
		 & \textbf{Freeform User Queries Scenario 1 (for Google, Hakia, DuckDuckGo)} \\ \hline
		1 & site:domain.name daily buprenorphine dosage $>$ 4mg \\ \hline
		2 & site:domain.name using bupe more than 4mg \\ \hline
		3 & site:domain.name buprenorphine dosage excess 4mg daily \\ \hline 
		4 & site:domain.name buprenorphine dosage more than 4mg daily \\ \hline
		5 & site:domain.name buprenorphine dosage over 4mg daily \\ \hline
		6 & site:domain.name (buprenorphine OR bup) dosage daily (``above 4mg'' OR ``over 4mg'' OR ``more than 4mg'')\\ \hline
	    \end{tabular}
\end{table}

\begin{figure*}[ht]
	\centering
		\caption{ Screenshot of search results from our hybrid information retrieval system for scenario1}\label{fig:predose}
		\includegraphics[scale=0.75]{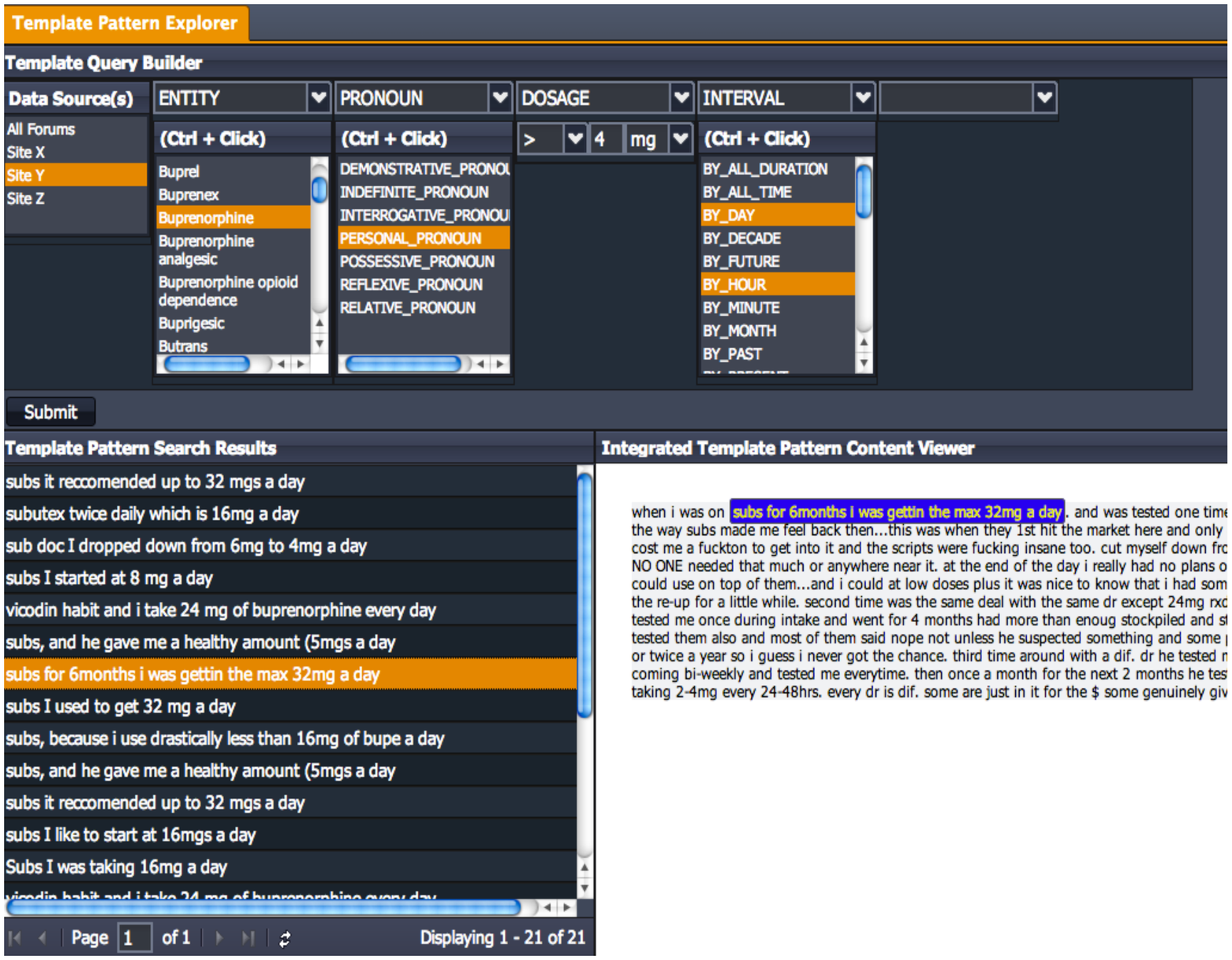}
\end{figure*}

Figure \ref{fig:predose} shows a screenshot of the results from our hybrid search system for this scenario. The selected web forum is indicated as Site Y under the Data Sources(s) panel (top left). As shown in the Template Query Builder interface (Figure \ref{fig:predose}, top right), to construct the user query for the information need, the evaluator must first select the template class {\footnotesize $\langle$ENTITY$\rangle$}, and then select the nonterminal `Buprenorphine' from the list. This concept is expanded according to the grammar production {\footnotesize $\langle  ENTITY \rangle$ $\rightarrow$ $L(C) \cup L_s(C) \cup L(R) \cup L_s(R)$}, which includes all slang terms and labels for all subclasses and individuals. The respondent then selected the template class {\footnotesize $\langle$PRONOUN$\rangle$} and selected the set of all {\footnotesize $\langle$PERSONAL PRONOUN$\rangle$}, which is interpreted according to a lexicon of pronouns. The evaluator then selected the template class {\footnotesize $\langle$DOSAGE$\rangle$}, which is interpreted according to the rules applied to the alphabet in the grammar (see \ref{grammarappendix}). Finally, the evaluator selected the template class {\footnotesize $\langle$INTERVAL$\rangle$} and then the nonterminals {\footnotesize $\langle$BY\_DAY$\rangle$} and {\footnotesize $\langle$BY\_HOUR$\rangle$}. 

The search results are shown in the Template Pattern Search Results Grid (Figure \ref{fig:predose}, bottom left) and the Integrated Template Pattern Content Viewer (Figure \ref{fig:predose}, bottom right). Among the search results, note first that all documents contained an amount greater than 4mg. Second, there were only 2 search results that contained the actual label `buprenorphine' in their annotation. This list of synonyms is as follows: subs--12, sub--2, Subutex--2, Suboxone--3, Buprenorphine--2. Both events can be attributed to the level of interpretation performed by the grammar. 

We note that a search result is considered if at least 4/6 evaluators agreed the result was relevant. This is reasonable because the significance of kappa scores across multiple users may diminish, but not necessarily indicate a major disagreement. Although Fleiss' kappa may be used instead, the results from this slight majority seems reasonable. And since each user query could be (and indeed was) different among the 6 evaluators, as shown in Table \ref{tab:uqs1}, relevance agreement for specific search results among the other sources (i.e., Google, Hakia and DuckDuckGo) is of little meaning. To compute the relevance of search results in the top 20, we considered agreement positionally, on the  relevance of each results across the separate lists. That is, among the evaluators how many agreed their first, second, third result and so on, were relevant.  Given the absence of a goal standard dataset for evaluation, as noted in \cite{Sure02, McCool05, BlancoHHMPTT13}, was deemed a reasonable compromise. 

\textbf{Scenario 2:}  The same evaluators repeated the evaluation for a different query scenario. This time, to find relevant documents that discuss: ``negative sentiments/experiences resulting from the use of Buprenorphine'' (Table \ref{tab:scenarios}, row 2). The selected free form queries used by the evaluators are listed in Table \ref{tab:uqs2}. The following specific user query was provided for use in our system: {\footnotesize $\langle$Buprenorphine$\rangle$} {\footnotesize $\langle$PERSONAL PRONOUN$\rangle$} {\footnotesize $\langle$NEGATIVE$\rangle$} for the template pattern {\footnotesize $\langle$ENTITY$\rangle$} {\footnotesize $\langle$PRONOUN$\rangle$} {\footnotesize $\langle$SENTIMENT$\rangle$}. 

\begin{table}
	\caption{User Queries for Scenario 2}
	\centering
 	\label{tab:uqs2}
	    \begin{tabular}{| c | p{7.5cm} |} \hline
		& \textbf{Freeform User Queries Scenario 2 (for Google, Hakia, DuckDuckGo)} \\ \hline
		1 & site:domain.name side effects for buprenorphine \\ \hline
		2 & site:domain.name side effect for buprenophine $|$ bup $|$ bupe $|$ bupes $|$ suboxone \\ \hline
		3 & site:domain.name buprenorphine bad experience \\ \hline 
		4 & site:domain.name ``buprenorphine'' ``horrible experience" \\ \hline
		5 & site:domain.name ``buprenorphine'' ``horrible\\ \hline
		6 & site:domain.name ``buprenorphine'' (``horrible'' OR ``side effect'') ``buprenorphine'' (``bad experience'' OR ``side effect'')\\ \hline
	    \end{tabular}
\end{table}

Figure \ref{fig:eval}, top right, shows that our system retrieved relevant 11/20 search results. This drop is likely because of the difficulty in interpreting the nonterminal {\footnotesize $\langle$NEGATIVE SENTIMENT$\rangle$}. Google retrieved 9/20 relevant results. However, we noticed that their search results did not contain sentiment-conveying keywords other than those specified by the user. Our system retrieved the following  sentiment expressions in the results (f*up--1, sh*t--2, weird--2, disappointed--1, f*ing weird--2, hated--1, rough--1, nauseous--1), which were not specified in the user query, but based on the interpretation of {\footnotesize $\langle$NEGATIVE$\rangle$} sentiment in the grammar (see \ref{grammarappendix}). Hakia and DuckDuckGo also performed less effectively, due to an inability to interpret {\footnotesize $\langle$NEGATIVE$\rangle$} sentiment as formulated in the user queries. 

\textbf{Scenario 3:}  Unlike the previous two scenarios, we asked 6 evaluators to search Google, Hakia and DuckDuckGo using the following specific search query: `site:domain.name buprenorphine dosage excess 4mg daily.' We performed this evaluation to assess the objective relevance judgments across the same search results. In this set, 3 evaluators were the same from Scenarios 1 and 2, and there were 3 new. Figure \ref{fig:predose} (bottom left) shows that this set of evaluators agreed that slightly more documents were relevant to the information need for the given query. A total of 18 out of 20 results were considered relevant in PREDOSE (an increase of 2), compared with 13 for Google (a decrease of 1). Within this set of 18 relevant search results, a few contained a mention of usage in the region of 32mg, which is considered an upper bound. We also observed that among the 13 relevant Google results, none of the highlighted amounts was greater than 4mg, but rather serendipitously contained greater amounts in the surrounding text. This reaffirms what is already known; Google does not interpret keywords to any significant degree, but rather performs keyword-based query expansion. DuckDuckGo showed a striking increase in the overall number of relevant results, increasing from 3/20 to 11/20. This suggests that the search query provided by our team for this evaluation was more effective in retrieving search results. Still however, it was observed that only few document snippets highlighted amounts that were greater than ``4mg.'' The relevant search results in Hakia also increased from 4/20 to 9/20 with only 3 highlighted amounts greater than ``4mg.''

\textbf{Scenario 4:}  Finally, the same procedure was repeated using the following specific query (`site:domain.name buprenorphine bad experience') provided by our team for the same information need from Scenario 2. That is, find relevant documents that discuss: ``negative sentiments/experiences resulting from the use of Buprenorphine.'' Figure \ref{fig:eval} (bottom right) shows that our system retrieved 11/20 relevant search results and again outperformed Google (7/20), DuckDuckGo (6/20) and Hakia (which notably did not retrieve any relevant results for the query).

\section{Related Work}\label{related}
In this section, we provide an overview of semantic search and hybrid information retrieval systems, which rely on semantic web technologies, and may be relevant to our problem. Semantic Web offers to create machine-processable representation of real-world concepts, whose meaning can be exploited for various tasks across heterogeneous information environments \cite{TBL2000}. Semantic search, as an application of semantic web technologies, is intended to enhance the retrieval of more accurate and high quality search results, when compared with traditional keyword-based search models and their various enhancements. An early realization of this idea has been the Semantic Content Organization and Retrieval Engine (SCORE) \cite{Hammond02, ShethBAHKW02, sheth2001system}, which uses both ontologies and rules (derived using regular expressions) for search. SCORE and its successor Semagix FREEDOM \cite{Sheth05enterpriseapplications, ShethAABWRHAAAK05}, were early platforms that integrated structured knowledge and additional intelligible constructs to support real-world and commercial knowledge-driven applications. 

In this work, we go beyond the functionality provided by  SCORE, by providing search support for: 1) classes of query elements modeled almost exclusively in lexicons (such as positive and negative sentiment expressions, and varying types and degrees of emotions), 2) information in lexicons, with only partial ontology representation, such as side effects, route-of-administration (ROA) and drug form, and 3) elements that belong to broad classes (including certain parts-of-speech), levels of intensity (high, low, average), and fuzzy interval references (past, present, future, etc). Moreover, we perform a deeper level of interpretation of certain expressions (such as `$>$4mg'), through a low-level CFG. We perform these tasks in addition to ontology-driven and rule-based search as was implemented in SCORE. 

Popov et. al. \cite{PopovKOMK04, KiryakovPTMO04} developed KIM that supports semantic annotation and search for entities and entity-patterns from the ontology that are also present in the corpus. The search is enhanced with support for lexical resources (such as currency, date, location, aliases, abbreviations etc) not typically represented in ontologies. To achieve this, Popov uses a modified pattern-matching grammar based on GATE , which recognizes relations in text, by gleaning entity associations from predicates in the ontology schema. While KIM is similar to our approach, we provide a more inclusive hybrid search system, which supports the retrieval of two additional types of data not considered by KIM: 1) data from lexicons, and 2) lexico-ontology concepts. Moreover, our system is more loosely coupled to accommodate query elements not in the ontology, while providing a broader range of pattern-based search through a top-level CFG. Additionally, we evaluate the relevance of search results for specific complex information needs in a domain specific setting. Popov et. al. \cite{PopovKOMK04, KiryakovPTMO04}  evaluate the precision and recall of annotations types (elements in our second-level grammar) rather than actual results of semantic search. In subsequent research, a search evaluation on television and radio news articles was conducted in \cite{DowmanTCP05} using KIM, based on ontology and keyword-based query interpretation, not the rules developed in the system. 

Guha et. al. \cite{GuhaMM03} presented a prototype semantic search system called TAP, that interprets keywords according to real-world concepts modeled in background knowledge. Various heuristics were used to find matching subgraphs for single keyword queries and  keyword pairs. The retrieved structured data was then rendered as an augmentation of search results from the document list, in a Google-style search interface. A key issue is that while the information gleaned from background knowledge may complement the search results in the document list, there is an assumption that query elements can be mapped to the ontology in the first place. This assumption will not always hold, as the authors themselves note, if ``the search term does not denote anything known to the Semantic Web, then we are not able to contribute anything to the search results.''

Thirunarayan and Immaneni in \cite{Immaneni09} also developed a hybrid query language to unify web of data and web of documents, This approach improves both: 1) information retrieval from Semantic Web through keyword-based search and 2) semantic search of hyperlinked web documents through the exploitation of inheritance hierarchy.  Their lucene-based SITAR (Semantic InformaTion Analysis and Retrieval) system provided enhanced retrieval from combined data sources such as AIFB SEAL. SITAR contains information about researchers that combines both structured and unstructured data. 

Lei et. al. \cite{LeiUM06} developed a semantic search engine called SemSearch, which is another ontology-driven system for keyword-based search over documents. However, SemSearch provides flexibility in query interpretation by also providing search support using Lucene, for keywords not present in the ontology. In the case of complex queries that contain multiple keywords, facts from the ontology are used as templates for query interpretation, similar to the approach in KIM \cite{PopovKOMK04}. 

Keyword-based semantic search systems offer a tradeoff between query expressiveness and accuracy of query interpretation, both of which affect the quality of the retrieved search results. Hence, there is a body of research focused on natural language query interfaces for semantic search \cite{LopezPM05, FernandezLSUVMC08, LopezMU06, FernandezCLVCM11, ValletFC05, CastellsFV07, Ruotsalo12}, to provide more query expressiveness. Lopez et. al.  developed AquaLog \cite{LopezPM05} and PowerAqua \cite{LopezMU06}, which translate natural language user queries into binary relational (triple) format, consistent with ontological representations. Fernandez et. al. \cite{FernandezLSUVMC08, FernandezCLVCM11} utilize PowerAqua for ontology-driven natural language query interpretation over documents. Queries expressed in natural language are translated into a formal representation using SPARQL. The document corpus is annotated based on entities, which populate the ontology knowledgebase of instances. In this way, the knowledgebase is a representation of the corpus, through the association of its annotations. While this approach is plausible (like SCORE), it requires that the corpus be represented in the ontology. The technique used for corpus annotation must necessarily be aware of the various types of query elements, extract them and represent them formally in the ontology. This is a challenging problem, as not all data types are suited for ontology representation, let alone querying using ontology query languages. 

\section{Discussion}\label{disc}
In this paper we present a knowledge-aware search system that utilizes a CFG, to enhance information retrieval in the domain of prescription drug abuse. The need for external resources to complement ontologies when addressing complex information needs is not isolated. In the development of Watson, the DeepQA research project \cite{FerrucciBCFGKLMNPSW10} at the IBM T. J. Watson Research Center, exploited a range of data sources, including ``unstructured data (e.g., typical web pages, blog posts) and semi-structured data (e.g., Wikipedia) to completely structured data (facts mined from the Web or [from] pre-existing databases)'' \cite{Ferrucci2008} for question answering (QA) in the \textit{Jeopardy! Challenge}. Our approach is consistent with this view, and other semantic search applications \cite{Hammond02, PopovKOMK04} capable of interpreting multifaceted queries involving ontological concepts and additional intelligible constructs. There continues to be a growing realization that to effectively address practical information retrieval problems, current knowledge models must be enhanced, by incorporating semantic enrichment modules capable of interpreting heterogeneous data. The implemented \textit{Template Pattern Explorer} in PREDOSE, is currently in use by epidemiologists at the Center for Interventions, Treatment and Addictions Research (CITAR) at the Wright State University. Moreover, the existing grammar is being extended for application to other social media and unstructured clinical notes, through existing (and proposed) project at Kno.e.sis. Some of these include: 1) clinical notes \cite{CameronBS12}, from which the grammar was inherited, 2) eDrugTrends -- \url{http://wiki.knoesis.org/index.php/EDrugTrends}, 3) biomaterials and materials science -- \url{http://wiki.knoesis.org/index.php/MaterialWays} and 4) knowledge acquisition from EMRs, which are project at Kno.e.sis (\url{http://knoesis.org}). 

While the approach outlined in this application shows early progress, there are several limitations. The first limitation is that the manual specification of the grammar for each application could be cumbersome and not scalable. General-purpose search engines perform well on the web due to the implementation of generic search algorithms. A method that can dynamically identify new and frequently occurring template patterns will have cross-domain implications. The 60 template patterns used in this study were created manually, based on information needs provided by domain experts. However, such resources may not be available in other domains.  While it is difficult to implement a generic algorithm that can effectively retrieve data for specific domains, a greater degree of automation in grammar composition is needed. Second, template-based query specification also requires considerable domain expertise and familiarity with the search application. A less restrictive query specification interface, such as one based on keywords or natural language, is under consideration. Another critical issue is the need for entity disambiguation to filter out spurious results. False positives impact the overall accuracy of the system and the relevance of the generated search results. While our earlier results, reported in \cite{CameronSDSDCACWF13}, showed that approximately 85\% of entities are correctly identified in our system, entities such as ``alcohol,'' ``cannibis'' and ``oxycontin'' are highly ambiguous. A context-aware methodology for entity disambiguation, such as that implemented by Mendes et. al. \cite{MendesJGB11} could be beneficial. Likewise, ranking search results, which is currently not provided, may be crucial in search scenarios where many search results exist. Popular concepts such as `methadone' and `heroin,' which occur with high frequency in the corpus can generate many search results, which may be overwhelming for domain experts to explore. In spite of these and many other limitations, this research is an early effort to develop a search system which addresses complex domain specific information retrieval in prescription drug abuse. We believe that overcoming the aforementioned limitations will only serve to improve the overall quality of the search system and make it more amenable to other domains. 

\section{Conclusion}\label{concl}
In this paper we presented a hybrid information retrieval system for domain specific information retrieval, applied to prescription drug abuse. The approach outlines a context-free grammar to specify the query language of expressions interpretable by the system. Consequently the hybrid system is capable of interpreting four types of query elements: 1) ontological knowledge, 2) concepts in lexicons, 3) concepts in lexicons with partial ontology representation (i.e., lexico-ontology) and 4) intelligible constructs defined exclusively through rules. The system uses template-based query specification to enable structured query composition and facilitate interpretation of domain specific query elements. In an evaluation against the contemporary semantic search system (Hakia), the crowd sourcing-based search system (DuckDuckGo) and the popular search engine Google, our system performed satisfactorily in retrieving relevant results for complex information needs. A live web application is currently available and in use by epidemiologists conducting research on emerging patterns and trends in prescription drug abuse using social media. The search system is also available online for option viewing -- \url{http://knoesis-hpco.cs.wright.edu/knowledge-aware-search}, together with an accompanying video demo \url{http://bit.ly/kasdemo}. 

\section{Author Contributions}
\noindent Delroy Cameron assisted in writing the manuscript and developed key aspects of the overall hybrid information retrieval system, and also contributed many ideas for the overall research. Amit P. Sheth established the interdisciplinary collaboration of the PREDOSE project, guided the development of the project, while providing ideas for its positioning within semantic search and also contributed to the writing. Krishnaprasad Thirunarayan developed key aspects of the context-free grammar and also provided crucial research ideas. Nishita Jaykumar, Gaurish Anand and Gary A. Smith assisted with many aspects of the system, including system implementation and evaluation, and the provision of various supporting online resources. 

\section{Conflict of Interest}
\noindent The authors would like to assert that there are no conflicts of interest regarding this manuscript.

\section{Acknowledgment}
\noindent The research is funded in part  by the National Institute on Drug abuse (NIDA) Grant No. 5R01DA039454-02 grant titled: ``Trending: Social media analysis to monitor cannabis and synthetic cannabinoid use'' and by Grant No. R21 DA030571-01A1. Any opinions, findings, conclusions or recommendations expressed in this material are those of the investigator(s) and do not necessarily reflect the views of the National Institutes of Health. We would also like to thank other members of the research team, including Raminta Daniulaityte, Drashti Dave, Revathy Krishnamurthy, Swapnil Soni and Kera Z. Watkins. 

\section{References}

\appendix
\section{Context-Free Grammar}\label{grammarappendix}
\noindent This appendix provides a partial listing of the context-free grammar (CFG) in \textit{Backus-Naur Form (BNF)}, which is used to interpret the query language $U$ of our hybrid information retrieval system $H$. Further details are available online: \url{http://wiki.knoesis.org/index.php/Knowledge-Aware-Search}

\subsection{Start Symbol}
\noindent The Start Symbol of the grammar is the nonterminal {\footnotesize $\langle$TEMPLATE PATTERN$\rangle$}. This start symbol supports productions containing sequences of {\footnotesize $\langle$TEMPLATE CLASS$\rangle$} nonterminals from the set $N^S$. There are 61 specific sequences of template classes supported by the start symbol\footnote{\url{http://wiki.knoesis.org/index.php/Knowledge-Aware-Search-Productions}}. To avoid listing these in detail the \textit{kleene start} operator is used in the first production below. The complete {list of productions} is available in the online supplementary materials (\url{http://wiki.knoesis.org/index.php/Knowledge-Aware-Search-Productions}). 

\begin{enumerate}
	\item {\footnotesize $\langle$TEMPLATE PATTERN$\rangle$ $\rightarrow$ $\langle$ TEMPLATE CLASS$\rangle$}$^\ast$
\end{enumerate}

\subsection{Template Classes}
\noindent There are 11 nonterminals in the set of template classes in $N^S$ that comprise the top-level grammar. 
\begin{enumerate}
	\setcounter{enumi}{1}
	\item {\footnotesize $\langle$TEMPLATE  CLASS$\rangle$ $\rightarrow$ 
		$\langle  INTERVAL \rangle |  
		\langle FREQUENCY \rangle$ $|$
		$\langle DOSAGE  \rangle$ $|
		\langle  ENTITY \rangle |
		\langle  ROA \rangle$ $|$
		$\langle  DRUGFORM \rangle$ $|$
		$\langle  SIDEEFFECT \rangle$ $|$
		$\langle  EMOTION \rangle$ $|$
		$\langle  PRONOUN \rangle$ $|$		
		$\langle  INTENSITY \rangle$ $|$	
		$\langle  SENTIMENT \rangle$}
\end{enumerate}

\subsection{Productions}
\noindent The set of nonterminals of in $G$ are shown in the following productions. \\

\noindent The \underline{{\footnotesize $\langle INTERVAL \rangle$}} template class is defined as follows:  

\begin{enumerate}
	\setcounter{enumi}{2}
	
	\item {\footnotesize $\langle  INTERVAL \rangle \rightarrow \langle DURATION\_PERIOD \rangle |
	\langle PERIOD\_DURATION\rangle$ $|
	\langle TIME\_PERIOD\rangle | 
	\langle PERIOD\_TIME\rangle |
	\langle AMOUNT\_TIME\_PERIOD\rangle$ $|
	\langle AMOUNT\_TIME\rangle |
	\langle PERIOD\_AMOUNT\_TIME\rangle$ $|
	\langle AMOUNT\_DURATION\_PERIOD\rangle |
	\langle AMOUNT\_DURATION\rangle$ $| 
	\langle PERIOD\_AMOUNT\_DURATION\rangle$}
	
	\item {\footnotesize $\langle DURATION\_PERIOD \rangle$ $\rightarrow$ 
		$\langle DURATION\_INDICATOR\rangle$$\langle RANGE\rangle$ $\langle PAST\_DETERMINER\rangle |
		\langle DURATION\_INDICATOR\rangle \langle RANGE\rangle$ $\langle PRESENT\_DETERMINER \rangle |
		\langle DURATION\_INDICATOR\rangle$ $\langle RANGE\rangle$ $\langle FUTURE\_DETERMINER\rangle$
	}
	
	\item {\footnotesize $\langle PERIOD\_DURATION \rangle$ $\rightarrow$ 
		$\langle PAST\_DETERMINER\rangle \langle RANGE\rangle$ $\langle DURATION\_INDICATOR\rangle |
		\langle PRESENT\_DETERMINER\rangle \langle RANGE\rangle$ $\langle DURATION\_INDICATOR\rangle |
		\langle FUTURE\_DETERMINER\rangle \langle RANGE\rangle$ $\langle DURATION\_INDICATOR\rangle$
	}
	
	\item {\footnotesize $\langle TIME\_PERIOD \rangle$ $\rightarrow$ 
		$\langle TIME\_INDICATOR\rangle \langle RANGE\rangle$ $\langle PAST\_DETERMINER\rangle | 
		\langle TIME\_INDICATOR\rangle \langle RANGE\rangle$ $\langle PRESENT\_DETERMINER\rangle |
		\langle TIME\_INDICATOR\rangle \langle RANGE\rangle$ $\langle FUTURE\_DETERMINER\rangle$
	}
	
	\item {\footnotesize $\langle PERIOD\_TIME \rangle$ $\rightarrow$ 
		$\langle PAST\_DETERMINER\rangle \langle RANGE\rangle$ $\langle TIME\_INDICATOR\rangle |
		\langle PRESENT\_DETERMINER\rangle \langle RANGE\rangle$ $\langle TIME\_INDICATOR\rangle |
		\langle FUTURE\_DETERMINER\rangle \langle RANGE\rangle$ $\langle TIME\_INDICATOR\rangle$ 
	}
	
	\item {\footnotesize $\langle AMOUNT\_TIME\_PERIOD \rangle$ $\rightarrow$ 
		$\langle AMOUNT\rangle \langle RANGE\rangle$ $\langle TIME\_PAST\_PERIOD\rangle | 
		\langle AMOUNT\rangle \langle RANGE\rangle$ $\langle TIME\_PRESENT\_PERIOD\rangle |
		\langle AMOUNT\rangle \langle RANGE\rangle$ $\langle TIME\_FUTURE\_PERIOD\rangle$
	}
	
	\item {\footnotesize $\langle AMOUNT\_TIME \rangle$ $\rightarrow$ 
		$\langle AMOUNT\rangle \langle RANGE\rangle$ $\langle DURATION\_PAST\_PERIOD\rangle | 
		\langle AMOUNT\rangle \langle RANGE\rangle$ $\langle DURATION\_PRESENT\_PERIOD\rangle | 
		\langle AMOUNT\rangle$ $\langle RANGE\rangle \langle DURATION\_FUTURE\_PERIOD\rangle$
	}
	
	\item {\footnotesize $\langle PERIOD\_AMOUNT\_TIME \rangle$ $\rightarrow$ 
		$\langle PAST\_DETERMINER \rangle$ $\langle RANGE \rangle$ $\langle AMOUNT\_TIME \rangle |
		\langle PRESENT\_DETERMINER \rangle \langle RANGE \rangle$ $\langle AMOUNT\_TIME \rangle | 
		\langle FUTURE\_DETERMINER \rangle \langle RANGE \rangle$ $\langle AMOUNT\_TIME \rangle$
	}
	
	\item {\footnotesize $\langle AMOUNT\_DURATION\_PERIOD \rangle$ $\rightarrow$ 
		$\langle AMOUNT\rangle \langle RANGE\rangle$ $\langle DURATION\_PAST\_PERIOD\rangle  |
		\langle AMOUNT\rangle \langle RANGE\rangle$ $\langle DURATION\_PRESENT\_PERIOD\rangle | 
		\langle AMOUNT\rangle \langle RANGE\rangle$ $\langle DURATION\_FUTURE\_PERIOD\rangle$
	}
	
	\item {\footnotesize $\langle AMOUNT\_DURATION\rangle \rightarrow  
		\langle AMOUNT\rangle \langle RANGE\rangle$ $\langle DURATION\_INDICATOR\rangle$
	}
	
	\item {\footnotesize $\langle PERIOD\_AMOUNT\_DURATION \rangle$ $\rightarrow$  
	$\langle PAST\_DETERMINER\rangle$ $\langle RANGE\rangle$ $\langle AMOUNT\_DURATION\rangle$ $|$
	$\langle PRESENT\_DETERMINER\rangle$ $\langle RANGE\rangle$ $\langle AMOUNT\_DURATION\rangle$ $|$
	$\langle FUTURE\_DETERMINER\rangle$ $\langle RANGE\rangle$ $\langle AMOUNT\_DURATION\rangle$}	

	\item {\footnotesize $\langle TIME\_INDICATOR \rangle$ $\rightarrow$ 
		$\langle HOUR\rangle$ $|$ 
		$\langle MINUTE \rangle$ $|$ 
		$\langle SECOND \rangle$
		} 
	
	\item {\footnotesize $\langle DURATION\_INDICATOR\ \rangle$ $\rightarrow$ 
		$\langle DECADE\rangle$ $|$
		$\langle YEAR\rangle$ $|$
		$\langle MONTH\rangle$ $|$
		$\langle WEEK\rangle$
	}
	
	\item {\footnotesize $\langle DECADE\rangle$ $\rightarrow$ \textbf{decade $|$ decades}} 
	\item {\footnotesize $\langle YEAR\rangle$ $\rightarrow$ \textbf{year $|$ years $|$ yr $|$ yrs $|$ annum}} 
	\item {\footnotesize $\langle MONTH\rangle$ $\rightarrow$ \textbf{month $|$ months $|$ mth $|$ mths $|$ mo}} 
	\item {\footnotesize $\langle WEEK\rangle$ $\rightarrow$ \textbf{week $|$ weeks $|$ wk $|$ wks}} 
	\item {\footnotesize $\langle DAY\rangle$ $\rightarrow$ \textbf{day $|$ night $|$ nite $|$ morning $|$ evening $|$ afternoon $|$ noon}} 
			
	\item {\footnotesize $\langle PERIOD \rangle \rightarrow \langle PAST\_DETERMINER\rangle |
	\langle PRESENT\_DETERMINER\rangle$ $|$ $\langle FUTURE\_DETERMINER\rangle$}
	
	\item {\footnotesize $\langle PAST\_DETERMINER \rangle$ $\rightarrow$ \textbf{ago $|$ prior $|$ previous $|$ since $|$ before $|$ \ldots}}
	\item {\footnotesize $\langle PRESENT\_DETERMINER \rangle$ $\rightarrow$ \textbf{now $|$ about $|$ around $|$ several $|$ \ldots}}
	\item {\footnotesize $\langle FUTURE\_DETERMINER \rangle$ $\rightarrow$ \textbf{next $|$ later $|$ after}}
			
	\item {\footnotesize $\langle SECOND\rangle$ $\rightarrow$ \textbf{second $|$ seconds $|$ sec $|$ secs}} 	
	\item {\footnotesize $\langle MINUTE\rangle$ $\rightarrow$ \textbf{minute $|$ minutes $|$ min $|$ mins}} 	
	\item {\footnotesize $\langle HOUR\rangle$ $\rightarrow$ \textbf{hour $|$ hours $|$ hr $|$ hrs}} 

	\item {\footnotesize $\langle NUMBER \rangle$} $\rightarrow \mathbb{N}$
	\item {\footnotesize $\langle NUMERIC\_AMOUNT \rangle$} $\rightarrow \mathbb{R}$
	\item {\footnotesize $\langle WORDED\_AMOUNT \rangle$} $\rightarrow$ {\footnotesize \textbf{one $|$ once $|$ two $|$ twice $|$ three $|$ thrice $|$ four $|$ five $|$ six $|$ seven $|$ eight $|$ nine $|$ ten $|$ eleven $|$ twelve $|$ thirteen $|$ fourteen $|$ fifteen $|$ sixteen $|$ seventeen $|$ eighteen $|$ nineteen $|$ twenty $|$ thirty $|$ forty $|$ fifty $|$ sixty $|$ seventy $|$ eighty $|$ nintey $|$ hundred}} 
	\item  {\footnotesize $\langle AMOUNT \rangle$ $\rightarrow$ $\langle NUMBER \rangle$ $|$ $\langle WORDED\_AMOUNT \rangle$}
	\item {\footnotesize $\langle RANGE \rangle$ $\rightarrow$  $[0 - \langle NUMBER\rangle$]}	
\end{enumerate}
\noindent The \underline{{\footnotesize $\langle FREQUENCY \rangle$}} template class is defined as follows:  
\begin{enumerate}
	\setcounter{enumi}{30}	
	\item {\footnotesize $\langle FREQUENCY \rangle \rightarrow
	\langle PER\_TIME\_INDICATOR\rangle$ $| 
	\langle PER\_DURATION\_INDICATOR\rangle$ $| 
	\langle AMOUNT\_FREQUENCY\_DURATION\rangle$ $| 
	\langle PERIOD\_FREQUENCY\_DURATION \rangle$ $| 
	\langle PERIOD\_FREQUENCY\_TIME  \rangle | 
	\langle AMOUNT\_FREQUENCY\_TIME \rangle$ $| 
	\langle AMOUNT\_PER\_TIME \rangle | 
	\langle AMOUNT\_PER\_DURATION \rangle$ $| 
	\langle FREQUENCY\_ITEM\rangle$
	}
		
		\item {\footnotesize $\langle AMOUNT\_FREQUENCY\_DURATION \rangle$ $\rightarrow$ 
		$\langle AMOUNT\_FREQUENCY\rangle$ 
		$\langle RANGE\rangle$ 
		$\langle DURATION\_INDICATOR\rangle$
		}
		
		\item {\footnotesize $\langle FREQUENCY\_DURATION \rangle$ $\rightarrow$ 
		$\langle FREQUENCY\_INDICATOR \rangle$  $|$
		$\langle RANGE\rangle$ $\langle DURATION\_INDICATOR \rangle$ 
		}
		
		\item {\footnotesize $\langle  FREQUENCY\_TIME \rangle$ $\rightarrow$ 
		$\langle FREQUENCY\_INDICATOR\rangle$ $\langle RANGE\rangle$ $\langle TIME\_INDICATOR\rangle$
		}

		\item {\footnotesize $\langle  PERIOD\_FREQUENCY\_DURATION \rangle$ $\rightarrow$ 
		$\langle PERIOD\_DETERMINER\rangle \langle RANGE\rangle \langle FREQUENCY\_INDICATOR\rangle$
		}
		
		\item {\footnotesize $\langle  PERIOD\_FREQUENCY\_TIME \rangle$ $\rightarrow$ 
		$\langle PERIOD\_DETERMINER\rangle$ $\langle RANGE\rangle$ 
		$\langle FREQUENCY\_TIME\rangle$
		}

		\item {\footnotesize $\langle AMOUNT\_FREQUENCY\_TIME \rangle$ $\rightarrow$ $\langle AMOUNT\rangle \langle RANGE\rangle \langle FREQUENCY\_TIME\rangle$}
		\item {\footnotesize $\langle  AMOUNT\_PER\_TIME \rangle$ $\rightarrow$ $\langle AMOUNT\rangle \langle RANGE\rangle \langle PER\_TIME\_INDICATOR\rangle$}
		\item {\footnotesize $\langle AMOUNT\_PER\_DURATION  \rangle$ $\rightarrow$ $\langle AMOUNT\rangle \langle RANGE\rangle \langle FREQUENCY\_TIME\rangle$}
	\item {\footnotesize $\langle FREQUENCY\_ITEM \rangle$ $\rightarrow$ \textbf{hourly $|$ daily $|$ weekly $|$ bi-weekly $|$ biweekly $|$ monthly $|$ yearly $|$ annually}}
		
	\item {\footnotesize $\langle PER\_INDICATOR\rangle$ $\rightarrow$ \textbf{per} $|$ / $|$ $\langle FREQUENCY\_INDICATOR\rangle$}	
	\item {\footnotesize $\langle PER\_SECOND\rangle$ $\rightarrow$ $\langle PER\_INDICATOR$ $\rangle$ $\langle RANGE\rangle$ $\langle SECOND\rangle$}
	\item {\footnotesize $\langle PER\_MINUTE\rangle$ $\rightarrow$ $\langle PER\_INDICATOR$ $\rangle$ $\langle RANGE\rangle$ $\langle MINUTE\rangle$}
	\item {\footnotesize $\langle PER\_HOUR\rangle$ $\rightarrow$ \textbf{hourly} $|$ $\langle PER\_INDICATOR$ $\rangle$ $\langle RANGE\rangle$ $\langle HOUR\rangle$}
	\item {\footnotesize $\langle PER\_DAY \rangle$ $\rightarrow$ \textbf{daily $|$ nightly $|$} $\langle PER\_INDICATOR\rangle$ $\langle RANGE\rangle$ $\langle DAY\rangle$} 
		\item {\footnotesize $\langle PER\_WEEK \rangle$ $\rightarrow$ \textbf{weekly $|$ bi-weekly $|$ biweekly $|$} $\langle PER\_INDICATOR\rangle$ $\langle RANGE\rangle$ $\langle WEEK\rangle$}	
		\item {\footnotesize $\langle PER\_MONTH \rangle \rightarrow  \textbf{monthly}  $|$ \langle PER\_INDICATOR\rangle \langle RANGE\rangle \langle MONTH\rangle$}	
		\item {\footnotesize $\langle PER\_YEAR \rangle$ $\rightarrow$ \textbf{yearly $|$ annually $|$} $\langle PER\_INDICATOR\rangle$ $\langle RANGE\rangle$ $\langle YEAR\rangle$}
		\item {\footnotesize $\langle PER\_DECADE \rangle$ $\rightarrow$ $\langle PER\_INDICATOR\rangle$ $\langle RANGE\rangle$ $\langle DECADE\rangle$}
		\item {\footnotesize $\langle  PER\_TIME\_INDICATOR \rangle$ $\rightarrow$ $\langle PER\_SECOND \rangle$ $|$ $\langle PER\_MINUTE \rangle$ $|$ $\langle PER\_HOUR \rangle$ $|$ $\langle PER\_DAY\rangle$ $|$ $\langle PER\_WEEK\rangle$ $|$ $\langle PER\_MONTH\rangle$ $|$ $\langle PER\_YEAR\rangle$ $|$ $\langle PER\_DECADE\rangle$}
		\item {\footnotesize $\langle  PER\_DURATION\_INDICATOR \rangle$ $\rightarrow$ $\langle PER\_INDICATOR\rangle$ $\langle RANGE\rangle$ $\langle DURATION\_INDICATOR \rangle$}
		\item {\footnotesize $\langle  AMOUNT\_PER\_TIME\_INDICATOR\rangle$ $\rightarrow \langle AMOUNT\rangle$ $\langle RANGE\rangle$ $\langle PER\_TIME\_INDICATOR\rangle$}
		\item {\footnotesize $\langle AMOUNT\_FREQUENCY \rangle \rightarrow \langle AMOUNT\rangle \langle RANGE\rangle$ $\langle FREQUENCY\_INDICATOR\rangle$}		
	\item {\footnotesize $\langle FREQUENCY\_INDICATOR\rangle$ $\rightarrow$ \textbf{times $|$ times a $|$ times an $|$ both times}}
	
\end{enumerate}

\noindent The \underline{{\footnotesize $\langle  DOSAGE \rangle$}} template class is as follows: 
\begin{enumerate}
	\setcounter{enumi}{55}	
	\item {\footnotesize $\langle DOSAGE  \rangle$ $\rightarrow$ $\langle NUMERIC\_AMOUNT\_UNIT\rangle$ $| \langle WORDED\_NUMERIC\_AMOUNT\_UNIT\rangle$}
\end{enumerate}

\noindent The \underline{{\footnotesize $\langle  ENTITY \rangle$}} template class is as follows: 
\begin{enumerate}
	\setcounter{enumi}{56}	
	\item {\footnotesize $\langle  ENTITY \rangle$ $\rightarrow$ $L(C) \cup L_s(C) \cup L(R) \cup L_s(R)$}
\end{enumerate}

\noindent The \underline{route-of-administration {\footnotesize $\langle  ROA \rangle$}} template class is defined as follows: 
\begin{enumerate}
	\setcounter{enumi}{57}	
	\item {\footnotesize $\langle  ROA \rangle \rightarrow 
		\langle ENTERAL\rangle | 
		\langle EPIDURAL\rangle |
		\langle INTRAARTERIAL\rangle$ $|	
		\langle INTRACARDIAC\rangle |
		\langle INTRACEREBRAL\rangle |	
		\langle INTRADERMAL\rangle$ $|
		\langle INTRAMUSCULAR\rangle |
		\langle INTRAVENOUS\rangle |
		\langle INHALATIONAL\rangle$ $|
		\langle INTRAPERITONEAL\rangle |
		\langle INTRATHECAL\rangle |
		\langle INTRAOSSEOUS$ $INFUSION\rangle |
		\langle NASAL\rangle |
		\langle PARENTERAL\rangle |
		\langle TRANSDERMAL\rangle$ $|
		\langle TRANSMUCOSAL\rangle$ $|$
		$\langle TOPICAL\rangle$ $|$
		$\langle SUBCUTANEOUS\rangle$
	}
	\item  {\footnotesize $\langle  INTRAPERITONEAL \rangle$ $\rightarrow$ \textbf{$\langle  INTRACEREBRAL \rangle$}}
	\item  {\footnotesize $\langle  INTRATHECAL \rangle$ $\rightarrow$ \textbf{$\langle  INTRACEREBRAL \rangle$}}
	\item  {\footnotesize $\langle INTRAOSSEOUS$ $INFUSION  \rangle$ $\rightarrow$ \textbf{$\langle  INTRACEREBRAL \rangle$}}
	\item  {\footnotesize $\langle  PARENTERAL \rangle$ $\rightarrow$ \textbf{$\langle  EPIDURAL \rangle$}}
	\item  {\footnotesize $\langle INTRAMUSCULAR  \rangle$ $\rightarrow$ \textbf{$\langle  EPIDURAL \rangle$ $|$ skin poppin}}
	\item  {\footnotesize $\langle  INTRAVENOUS \rangle$ $\rightarrow$ \textbf{$\langle  INTRAARTERIAL \rangle$}}
	\item  {\footnotesize $\langle TOPICAL  \rangle$ $\rightarrow$ \textbf{$\langle TRANSDERMAL  \rangle$}}
	\item  {\footnotesize $\langle  SUBCUTANEOUS \rangle$ $\rightarrow$ \textbf{$\langle  INTRACEREBRAL \rangle$}}
	\item  {\footnotesize $\langle INTRADERMAL  \rangle$ $\rightarrow$ \textbf{$\langle  INTRAARTERIAL \rangle$ $|$ sniff $|$ snort $|$ snorting $|$ bumping $|$ railing $|$ doozing}}
	\item  {\footnotesize $\langle ENTERAL  \rangle$ $\rightarrow$ \textbf{ate $|$ chewing $|$ drink $|$ eat $|$ insufflate $|$ plug $|$ plugged $|$ smoke $|$ smoked $|$ sniff $|$ snort $|$ \ldots}}
	\item  {\footnotesize $\langle EPIDURAL  \rangle$ $\rightarrow$ \textbf{inject $|$ injected $|$ injection}}
	\item  {\footnotesize $\langle  INTRAARTERIAL \rangle$ $\rightarrow$ \textbf{IV $|$ IVed $|$ IVing $|$inject $|$ injected $|$ \ldots}}
	\item  {\footnotesize $\langle INTRACARDIAC  \rangle$ $\rightarrow$ $\langle EPIDURAL  \rangle$}
	\item  {\footnotesize $\langle  INTRACEREBRAL \rangle$ $\rightarrow$ $\langle EPIDURAL  \rangle$}
	\item  {\footnotesize $\langle  INHALATIONAL \rangle$ $\rightarrow$ \textbf{smoke $|$ smokes $|$ smoked $|$ smoking $|$ sniff $|$ sniffed $|$ sniffing $|$ snort $|$ snorted $|$ snorting $|$ bumping $|$ railing $|$ doozing}}
	\item  {\footnotesize $\langle  NASAL \rangle$ $\rightarrow$ \textbf{sniff $|$ snort $|$ snorting $|$ bumping $|$ railing $|$ doozing}}
	\item  {\footnotesize $\langle TRANSDERMAL  \rangle$ $\rightarrow$ \textbf{patch $|$ patches}}
	\item  {\footnotesize $\langle  TRANSMUCOSAL \rangle$ $\rightarrow$ \textbf{snort $|$ snorted $|$ snorting $|$ sniff $|$ sniffed $|$ sniffing $|$ bumping $|$ railing $|$ doozing}}	

\end{enumerate}

\noindent The \underline{{\footnotesize $\langle  DRUGFORM \rangle$}} template class is defined as follows: 		
\begin{enumerate}
	\setcounter{enumi}{76}
	\item {\footnotesize $\langle  DRUGFORM \rangle$ $\rightarrow$ $\langle LIQUID\rangle$ $|$ $\langle SOLID\rangle$}
	\item {\footnotesize $\langle LIQUID  \rangle$ $\rightarrow$ \textbf{syrups $|$ elixirs $|$ suspensions $|$ ointment $|$\ldots}}
	\item {\footnotesize $\langle SOLID  \rangle$ $\rightarrow$ \textbf{powder $|$ tablet $|$  tablets $|$ tab $|$ tabs $|$ pill $|$ \ldots}}
\end{enumerate}

\noindent The \underline{{\footnotesize $\langle  SIDEEFFECT \rangle$}} template class is defined as follows: 		
\begin{enumerate}
	\setcounter{enumi}{79}	
	\item {\footnotesize $\langle  SIDEEFFECT \rangle$ $\rightarrow$ $\langle MILD\rangle$ $|$  $\langle MODERATE\rangle$ $|$  $\langle SEVERE\rangle$}
	\item  {\footnotesize $\langle MILD  \rangle$ $\rightarrow$ \textbf{bruising $|$ itching $|$ itching of skin $|$ tingling $|$ \ldots}}
	\item  {\footnotesize $\langle MODERATE  \rangle$ $\rightarrow$ \textbf{blisters $|$ blistering $|$ skin blisters that are itchy $|$ skin blisters that are painful $|$ skin discoloration $|$ \ldots}}
	\item  {\footnotesize $\langle SEVERE  \rangle$ $\rightarrow$ \textbf{abnormal heartbeat $|$ bone pain $|$ chest pain $|$ chest discomfort $|$ chest tightness $|$ chills $|$ coma $|$ \ldots}}
\end{enumerate}
	
\noindent The \underline{{\footnotesize $\langle  EMOTION \rangle$}} template class is defined as follows: 		
\begin{enumerate}
	\setcounter{enumi}{83}	
	\item {\footnotesize $\langle  EMOTION \rangle$ $\rightarrow$ 
		$\langle AFFECTION \rangle$ $|$ 
		$\langle LUST \rangle$ $|$ 
		$\langle LONGING \rangle$ $|$ 
		$\langle CHEERFULNESS \rangle$ $|$ 
		$\langle ZEST \rangle$ $|$ 
		$\langle CONTENTMENT \rangle$ $|$ 
		$\langle PRIDE \rangle$ $|$ 
		$\langle OPTIMISM \rangle$ $|$ 
		$\langle ENTHRALLMENT \rangle$ $|$ 
		$\langle RELIEF \rangle$ $|$ 
		$\langle SURPRISE \rangle$ $|$ 
		$\langle IRRITATION \rangle$ $|$ 
		$\langle EXASPERATION \rangle$ $|$
		$\langle RAGE \rangle$ $|$
		$\langle DISGUST \rangle$ $|$
		$\langle ENVY \rangle$ $|$
		$\langle TORMENT \rangle$ $|$
		$\langle SUFFERING \rangle$ $|$
		$\langle DEPRESSION \rangle$ $|$
		$\langle DISAPPOINTMENT \rangle$ $|$
		$\langle SHAME \rangle$ $|$
		$\langle NEGLECT \rangle$ $|$
		$\langle SYMPATHY \rangle$ $|$
		$\langle HORROR \rangle$ $|$
		$\langle CONFUSE \rangle$ $|$
		$\langle DISCONTENTMENT \rangle$ $|$
		$\langle EMBARRASSMENT \rangle$ $|$
		$\langle FORGIVENESS \rangle$ $|$
		$\langle THANKFULNESS \rangle$ $|$
		$\langle BLAME \rangle$ $|$
		$\langle NERVOUSNESS \rangle$ $|$
		$\langle LOVE \rangle$ $|$
		$\langle JOY \rangle$ $|$		
		$\langle ANGER \rangle$ $|$
		$\langle SADNESS \rangle$ $|$
		$\langle FEAR \rangle$
		}

	\item  {\footnotesize $\langle  LOVE \rangle$ $\rightarrow$ \textbf{$\langle  AFFECTION \rangle$ $\langle  LUST \rangle$ $\langle  LONGING \rangle$}}
	\item  {\footnotesize $\langle  JOY \rangle$ $\rightarrow$ \textbf{$\langle  CHEERFULNESS \rangle$ $\langle  ZEST \rangle$  $\langle  CONTENTMENT \rangle$ $\langle  PRIDE \rangle$ $\langle  OPTIMISM \rangle$ $\langle  ENTHRALLMENT \rangle$ $\langle  RELIEF \rangle$}}
	\item  {\footnotesize $\langle  ANGER \rangle$ $\rightarrow$ \textbf{$\langle  IRRITATION \rangle$ $\langle  EXASPERATION \rangle$ $\langle  RAGE \rangle$ $\langle  DISGUST \rangle$ $\langle  ENVY \rangle$ $\langle  TORMENT \rangle$}}
	\item  {\footnotesize $\langle SADNESS  \rangle$ $\rightarrow$ \textbf{$\langle SUFFERING  \rangle | \langle  DEPRESSION \rangle | \langle  DISAPPOINTMENT \rangle$ $| \langle  SHAME \rangle | \langle   NEGLECT\rangle | \langle   SYMPATHY\rangle$}}																
	\item  {\footnotesize $\langle FEAR  \rangle$ $\rightarrow$ \textbf{$\langle  HORROR \rangle$ $|$ $\langle  NERVOUSNESS \rangle$}}	
	
	\item  {\footnotesize $\langle  AFFECTION \rangle$ $\rightarrow$ \textbf{adoration $|$ affection $|$ love $|$ fondness $|$ liking $|$ attraction $|$ caring $|$ \ldots}}
	\item  {\footnotesize $\langle  LUST \rangle$ $\rightarrow$ \textbf{arousal $|$ desire $|$ lust $|$ lusting $|$ passion $|$ infatuation}}
	\item  {\footnotesize $\langle  LONGING \rangle$ $\rightarrow$ \textbf{longing}}
	\item  {\footnotesize $\langle  CHEERFULNESS \rangle$ $\rightarrow$ \textbf{amused $|$ amusement $|$ bliss $|$ blithe $|$ \ldots}}	
	\item  {\footnotesize $\langle  ZEST \rangle$ $\rightarrow$ \textbf{enthusiasm $|$ zeal $|$ zest $|$ excited $|$ exciting $|$ excitement $|$ thrill $|$ thrilling $|$ exhilaration}}
	\item  {\footnotesize $\langle  CONTENTMENT \rangle$ $\rightarrow$ \textbf{contented $|$ contentedness $|$ contentment $|$ pleasure $|$ satisfied $|$ satisfaction $|$ gratified $|$ gratification}}		
	\item  {\footnotesize $\langle PRIDE  \rangle$ $\rightarrow$ \textbf{pride $|$ proud $|$ prideful $|$ pridefulness $|$ triumph}}		
	\item  {\footnotesize $\langle  OPTIMISM \rangle$ $\rightarrow$ \textbf{eagerness $|$ expecting $|$ hope $|$ hopeful $|$ hoping $|$ hopefulness $|$ optimistic $|$ optimism}}		
	\item  {\footnotesize $\langle  ENTHRALLMENT \rangle$ $\rightarrow$ \textbf{enthrallment $|$ enthrall $|$ rapture}}		
	\item  {\footnotesize $\langle  RELIEF \rangle$ $\rightarrow$ \textbf{relief $|$ ease $|$ relaxation $|$ alleviation}}		
	\item  {\footnotesize $\langle  SURPRISE \rangle$ $\rightarrow$ \textbf{amazement $|$ amazed $|$ surprise $|$ surprised $|$ surprising $|$ astonished $|$ astonishment $|$ astounded $|$ unexpected}}		
	\item  {\footnotesize $\langle   IRRITATION\rangle$ $\rightarrow$ \textbf{aggravation $|$ irritation $|$ irritated $|$ irritating $|$ agitation $|$ annoyed $|$ annoyance $|$ disturbing $|$ grouchiness $|$ grumpiness}}		
	\item  {\footnotesize $\langle  EXASPERATION \rangle$ $\rightarrow$ \textbf{exasperation $|$ frustration}}				
	\item  {\footnotesize $\langle  RAGE \rangle$ $\rightarrow$ \textbf{anger $|$ rage $|$ outrage $|$ fury $|$ wrath $|$ hostility $|$ \ldots}}		
	\item  {\footnotesize $\langle  DISGUST \rangle$ $\rightarrow$ \textbf{disgust $|$ revulsion $|$ contempt $|$ disgusting $|$ disgusted}}		
	\item  {\footnotesize $\langle   ENVY\rangle$ $\rightarrow$ \textbf{envy $|$ jealousy $|$ jealous $|$ envying}}					
	\item  {\footnotesize $\langle  TORMENT \rangle$ $\rightarrow$ \textbf{torment $|$ tormented}}					
	\item  {\footnotesize $\langle SUFFERING  \rangle$ $\rightarrow$ \textbf{aggravation $|$ irritation $|$ irritated $|$ irritating $|$ \ldots}}					
	\item  {\footnotesize $\langle  DEPRESSION \rangle$ $\rightarrow$ \textbf{depressed $|$ depression $|$ depressing $|$ cheerless $|$ despair $|$ despairing $|$ \ldots}}				
	\item  {\footnotesize $\langle  DISAPPOINTMENT \rangle$ $\rightarrow$ \textbf{dismay $|$ disappointment $|$ disappointed $|$ disappointing $|$ displeasure $|$ letdown}}										
	\item  {\footnotesize $\langle  SHAME \rangle$ $\rightarrow$ \textbf{ashamed $|$ shame $|$ regret $|$ regretful $|$ regretting $|$ remorseful $|$ guilt $|$ remorse $|$ guilty $|$ \ldots}}										
	\item  {\footnotesize $\langle   NEGLECT\rangle$ $\rightarrow$ \textbf{alienation $|$ isolation $|$ neglect $|$ loneliness $|$ \ldots}}				
	\item  {\footnotesize $\langle   SYMPATHY\rangle$ $\rightarrow$ \textbf{pity $|$ sympathy $|$ compassion $|$ compassionate $|$ \ldots}}		
	\item  {\footnotesize $\langle  HORROR \rangle$ $\rightarrow$ \textbf{alarm $|$ shock $|$ hysteria $|$ mortification $|$ \ldots}}					
	\item  {\footnotesize $\langle   CONFUSE\rangle$ $\rightarrow$ \textbf{confused $|$ confusing $|$ confusion $|$ confuse}}	
	\item  {\footnotesize $\langle  DISCONTENTMENT \rangle$ $\rightarrow$ \textbf{discontent $|$ discontented $|$ \ldots}}	
	\item  {\footnotesize $\langle EMBARRASSMENT  \rangle$ $\rightarrow$ \textbf{embarrassment $|$ embarrass $|$ \ldots}}	
	\item  {\footnotesize $\langle   FORGIVENESS\rangle$ $\rightarrow$ \textbf{forgiveness $|$ forgive $|$ pardon $|$ forgiving}}	
	\item  {\footnotesize $\langle  THANKFULNESS \rangle$ $\rightarrow$ \textbf{thankfulness $|$ thankful $|$ appreciation $|$ \ldots}}	
	\item  {\footnotesize $\langle  BLAME \rangle$ $\rightarrow$ \textbf{blame $|$ blamed $|$ blaming $|$ \ldots}}	
	\item  {\footnotesize $\langle  NERVOUSNESS \rangle$ $\rightarrow$ \textbf{anxiety $|$ nervousness $|$ tenseness $|$ \ldots}}	

\end{enumerate}

\noindent The \underline{{\footnotesize $\langle  PRONOUN \rangle$}} template class is defined as follows: 
\begin{enumerate}
	\setcounter{enumi}{120}			
	\item {\footnotesize $\langle  PRONOUN \rangle$ $\rightarrow$ 
		$\langle DEMONSTRATIVE\_PRONOUN\rangle$ $|
		\langle PERSONAL\_PRONOUN\rangle | 	
		\langle POSSESSIVE\_PRONOUN\rangle$ $|		
		\langle REFLEXIVE\_PRONOUN\rangle |	
		\langle RELATIVE\_PRONOUN\rangle$ $|
		\langle INDEFINITE\_PRONOUN\rangle |
		\langle INTERROGATIVE\_PRONOUN\rangle$
	}
	\item  {\footnotesize $\langle DEMONSTRATIVE\_PRONOUN\rangle$ $\rightarrow$ \textbf{this $|$ that $|$ these $|$ those $|$ \ldots}}
	\item  {\footnotesize $\langle PERSONAL\_PRONOUN\rangle$ $\rightarrow$ \textbf{i $|$ me $|$ you $|$ she $|$ her $|$ he $|$ \ldots}}
	\item  {\footnotesize $\langle POSSESSIVE\_PRONOUN\rangle$ $\rightarrow$ \textbf{my $|$ our $|$ ours $|$ your $|$ yours $|$ \ldots}}				
	\item  {\footnotesize $\langle REFLEXIVE\_PRONOUN\rangle$ $\rightarrow$ \textbf{myself $|$ ourselves $|$ yourself $|$ \ldots}}				
	\item  {\footnotesize $\langle RELATIVE\_PRONOUN\rangle$ $\rightarrow$ \textbf{that $|$ which $|$ who $|$ whom $|$ whose $|$ \ldots}}				
	\item  {\footnotesize $\langle INDEFINITE\_PRONOUN\rangle$ $\rightarrow$ \textbf{anybody $|$ anyone $|$ anything $|$  \ldots}}				
	\item  {\footnotesize $\langle INTERROGATIVE\_PRONOUN\rangle$ $\rightarrow$ \textbf{what $|$ who $|$ which $|$ whom $|$ \ldots}}	
\end{enumerate}
		
\noindent The \underline{{\footnotesize $\langle INTENSITY \rangle$}} template class is defined as follows: 
\begin{enumerate}
	\setcounter{enumi}{128}			
	\item {\footnotesize $\langle  INTENSITY \rangle$ $\rightarrow$ $\langle LOW\rangle$ $|$ $\langle AVERAGE\rangle$ $|$ $\langle HIGH\rangle$}
	\item {\footnotesize $\langle  LOW \rangle$ $\rightarrow$ \textbf{low $|$ very low $|$ lower $|$ lower than $|$ lowest $|$ \ldots}}
	\item {\footnotesize $\langle  AVERAGE \rangle$ $\rightarrow$ \textbf{average $|$ ideal $|$ \ldots}}
	\item {\footnotesize $\langle  HIGH \rangle$ $\rightarrow$ \textbf{high $|$ very high $|$ higher $|$ highest $|$ large $|$ \ldots}}
\end{enumerate}

\noindent The \underline{{\footnotesize $\langle  SENTIMENT \rangle$}} template class is defined as follows: 
\begin{enumerate}
	\setcounter{enumi}{132}		
	\item {\footnotesize $\langle  SENTIMENT \rangle$ $\rightarrow$ $\langle POSITIVE\rangle$ $|$  $\langle NEGATIVE\rangle$  $|$  $\langle NEUTRAL \rangle$}	
	\item  {\footnotesize $\langle  POSITIVE \rangle$ $\rightarrow$ \textbf{Im glad $|$  luckily $|$ awesome $|$ benefit $|$ best choices $|$ best for me $|$ best $|$ \ldots}}
	\item  {\footnotesize $\langle  NEGATIVE \rangle$ $\rightarrow$ \textbf{big f*cking mistake $|$ threw up $|$ It was bad $|$ Its really rough $|$ \ldots}}
	\item  {\footnotesize $\langle  NEUTRAL \rangle$ $\rightarrow$ \textbf{hope $|$ longer $|$ well $|$ as well $|$ \ldots}} 
	\item {\footnotesize $\rightarrow$ $\langle$ ENTITY$\rangle$ $\langle$ INTERVAL $\rangle$ $\langle$ SIDEEFFECT $\rangle$}	
\end{enumerate}

\noindent The set of contextual compilation defines additional semantics for common constructs, which can be easily extended and reused across different domains.  
\begin{enumerate}
	\setcounter{enumi}{136}
	\item  {\footnotesize$\langle greaterThanOp \rangle$ $\rightarrow$ \textbf{$>$ $|$ greater than $|$ more than $|$ above $|$ in excess of $|$ slightly above $|$ little more$|$ bit more $|$ slightly more $|$ high} $|$\ldots} 
	\item  {\footnotesize $\langle lessThanOp \rangle$ $\rightarrow$ \textbf{$<$ $|$ less than $|$ lower than $|$ below $|$ in lack of $|$ slightly below $|$ little less $|$ bit less $|$ slightly less} $|$\ldots}
	\item  {\footnotesize $\langle equalToOp \rangle$  $\rightarrow$ \textbf{$=$ $|$ exactly $|$ precisely} $|$ \ldots}
	\item  {\footnotesize $\langle greaterThanEqualToOp \rangle$ $\rightarrow$ \textbf{$>=$ $|$ greater than $|$ greater than or equal to $|$ more than $|$ above $|$ in excess of $|$ slightly above $|$ little more $|$ bit more $|$ slightly more $|$ exactly $|$ precisely $|$ high} $|$ \ldots}
	\item  {\footnotesize $\langle lessThanEqualToOp \rangle$ $\rightarrow$ \textbf{$<=$  $|$ less than $|$ less than or equal to $|$  lower than $|$ below $|$ in lack of $|$ slightly below $|$ little less $|$ bit less $|$ slightly less $|$ exactly $|$ precisely} $|$ \ldots}	
\end{enumerate}

\end{document}